\documentclass[10pt,aps,prl,twocolumn,superscriptaddress,longbibliography]{revtex4-2}

\usepackage[dvipsnames,x11names]{xcolor}\colorlet{red}{black}
\usepackage[normalem]{ulem}
\usepackage{graphicx}
\usepackage{listings}

\usepackage{stmaryrd}
\SetSymbolFont{stmry}{bold}{U}{stmry}{m}{n}

\usepackage{amssymb}
\usepackage{amsmath}
\usepackage{gensymb}
\usepackage{textcomp} 
\usepackage{bbold}
\usepackage{bm}
\usepackage{placeins}
\usepackage{lmodern}

\usepackage[
    colorlinks=true,
    linkcolor=blue,
    citecolor=blue,
    urlcolor=blue
]{hyperref}
\newcommand{\Tr}{{\rm Tr}}

\newcommand{\Sa}{\bm{\sigma}^{{\rm a}}}

\newcommand{\rhoIN}{\rho_{\text{\tiny\textrm{IN}}}}
\newcommand{\phiIN}{\phi_{\text{\tiny\textrm{IN}}}}

\begin{document}
\title{Controlling Turbulent Flows in Compressible Active \textcolor{red}{Liquid Crystals}}

\author{Dimitrios Krommydas}
\email{dimitris@ucsb.edu}
\affiliation{Department of Physics, University of California Santa Barbara,
CA 93106, USA}

\author{Paarth Gulati}
\affiliation{Department of Physics, Emory University, Atlanta, GA, USA}
\affiliation{Initiative in Theory and Modeling of Living Systems, Emory University, Atlanta, GA, USA}

\author{Aparna Baskaran}
\affiliation{Department of Physics, Brandeis University, 415 South Street, Waltham, Massachusetts 02453, USA}

\author{M. Cristina Marchetti}
\affiliation{Department of Physics, University of California Santa Barbara, CA 93106, USA}
\affiliation{Interdisciplinary Program in Quantitative Biosciences, University of California Santa Barbara, Santa Barbara, CA 93106, USA}

\date{\today}

\begin{abstract}
{Motivated by experiments on light-patterned, quasi-2D active suspensions that exhibit large density variations, we develop a continuum theory of \textit{compressible} active nematics---suspensions of apolar rods whose orientation is invariant under $\pi$ rotations. Under spatially patterned activity, the extensile isotropic active pressure expels material from high-activity regions and accumulates it in low-activity ones; an exactly solvable scalar reduction shows that the resulting density contrast is governed by a single dimensionless parameter, linear in the compressibility. Using compressibility as a tuning knob, we then steer active turbulence from high- to low-activity regions and, at sharp activity interfaces, stabilize an analytically tractable dynamical steady state: a one-dimensional chain of vortices held by soft, activity-induced confinement. Our results establish compressibility as a control parameter for density variations and turbulent flow in active nematic suspensions.}

\end{abstract}

\maketitle
Spatiotemporal patterning of activity drives mechanobiological functions across living systems, from subcellular cytoskeletal assemblies to tissue and organ morphogenesis~\cite{saadaoui2020tensile, martin2010integration, behrndt2012forces, etournay2015interplay, mitchell2022visceral, nerurkar2019molecular}. Despite the underlying biochemical complexity, these processes ultimately rely on the generation and transmission of mechanical stresses to transport, reorganize, and position material. A defining feature of such processes is that activity and density vary together in space: where active stresses concentrate material, they reshape the very density field that sets their magnitude. Understanding how spatially varying activity organizes flow \emph{and} density---and the physical constraints on doing so---remains a central challenge~\cite{marchetti2013hydrodynamics, prost2015active, ramaswamy2010mechanics}.

Experiments now realize precisely this regime. In  reconstituted systems, model microorganisms, and synthetic active matter, light patterns control the structure, dynamics, and organization of active particles~\cite{palacci2013living, dellarcipe2018active, vizsnyiczai2017light, frangipane2018dynamic, arlt2018painting, aubret2018targeted}, and optogenetic tools~\cite{ross2019controlling} have made activity gradients directly programmable. In microtubule--kinesin suspensions with light-activatable motors, activity is set by light intensity, enabling large and controllable activity gradients~\cite{zarei2023light,lemma2023spatiotemporal,ross2019controlling, zhang2021spatiotemporal, chen2022selfmixing, nishiyama2025closedloop,yang2025dynamic}. These gradients drive correspondingly large density variations: active stresses push material between high-activity and low-activity regions, so that density becomes spatially structured and can no longer be treated as constant~\cite{marchetti2013hydrodynamics, prost2015active, marenduzzo2007steady, edwards2009spontaneous, simha2002hydrodynamic, doostmohammadi2018active}. This is a generic consequence of the underlying physics---active fluids such as microtubule--kinesin mixtures, actomyosin networks, and bacterial suspensions are two-fluid systems whose active constituents form a \emph{compressible} medium~\cite{sanchez2012spontaneous, pedley1992hydrodynamic, edwards2009spontaneous}.

Most existing continuum theories have treated compressibility and hydrodynamic flow in separate limits. Dry active matter models are compressible but neglect momentum-conserving flows~\cite{maryshev2019dry,toner1995long, cates2015motility, geyer2018sounds}, while wet active nematic theories incorporate hydrodynamic flows but typically assume incompressibility, suppressing the density variations that patterned-activity experiments make central~\cite{giomi2014defect,giomi2015geometry, doostmohammadi2018active, marenduzzo2007steady, shankar2022topological, ruske2022activity,partovifard2024controlling,
schimming2025turbulence}. A framework that treats flow and large density variations on the same footing---and uses one to control the other---has been largely absent, a gap made acute by recent  experiments in which density fluctuations are a dominant driver of the dynamics~\cite{Cheng2026}.

In this Letter we close this gap by developing a hydrodynamic theory of \textit{compressible} active nematics that consistently incorporates both flow and density variations, and use it to \emph{control} turbulent flows. The theory unifies existing descriptions---reducing to standard wet active nematics in the incompressible limit~\cite{simha2002hydrodynamic, thampi2013velocity, hemingway2016correlation, mukherjee2026turbulence} and to dry active matter models in the overdamped limit~\cite{dey2012spatial, liebchen2017phoretic, mishra2010fluctuations}---while retaining compressibility as an explicit, experimentally accessible parameter. Exploiting this, we show that compressibility sets the density contrast under patterned activity through the active P\'eclet number, and thereby steers active turbulence between high- and low-activity regions, culminating in a novel dynamical steady state: a one-dimensional chain of vortices held by soft, activity-induced confinement.

\begin{figure}[!t]
  \centering
\includegraphics[width=\columnwidth]{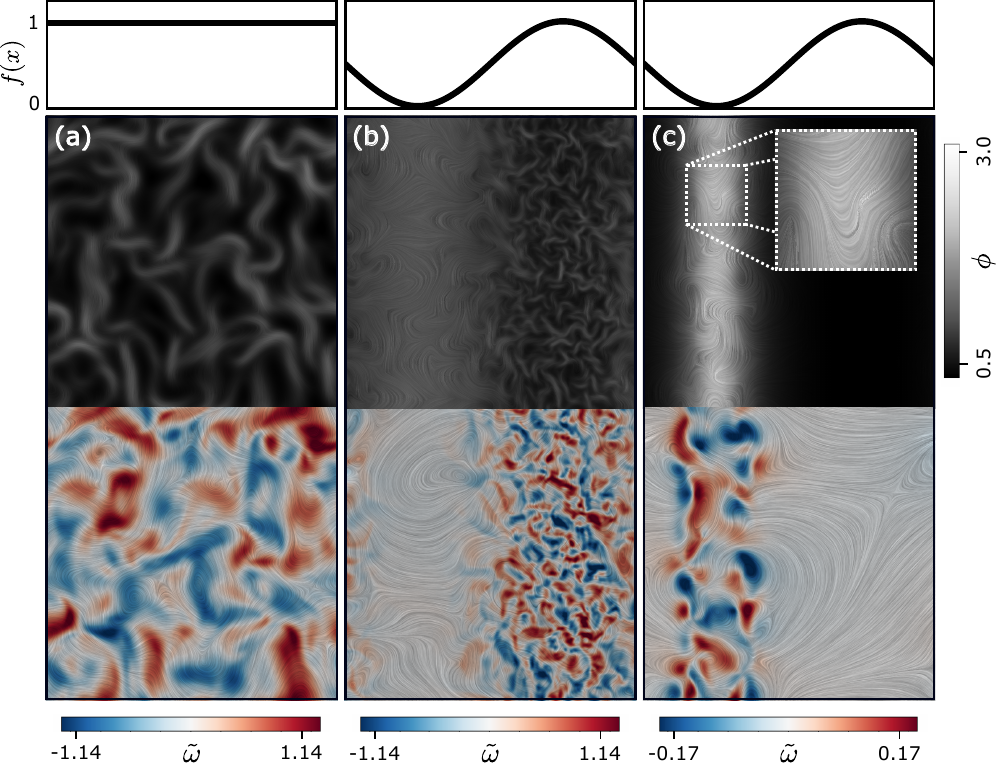}
  \caption{\textbf{Turbulent 
  patterns in compressible active nematics.}
  The black curves above the panels show the imposed dimensionless activity profiles $f(x)$.
  \textbf{(a--c)} 
  Top: density and director patterns, with white denoting high density.
  Bottom: vorticity (color) and flow streamlines.
  \textbf{(a)} Active turbulent flows for uniform activity, showing density inhomogeneities, for $\kappa=1.43$.
  \textbf{(b)} Sinusoidal activity profile 
  for $\kappa=1.43$, showing density accumulation in the low-activity region and depletion in the high-activity region, where turbulent flows localize.
  \textbf{(c)} Same as (b) for $\kappa=20$, showing a much larger density difference between the high- and low-activity regions. Note that flows are now localized in the low-activity region.
  In all panels, the vorticity $\tilde{\omega}$ is normalized by the maximum vorticity obtained from the numerical solution of the incompressible analog of Eqs.~\eqref{eq:mass}--\eqref{eq:nematic}, with all other numerical parameters unchanged.
  }
  \label{fig1:result_summary}
\end{figure}

\begin{figure}[!t]
  \centering
\includegraphics[width=0.95\columnwidth]{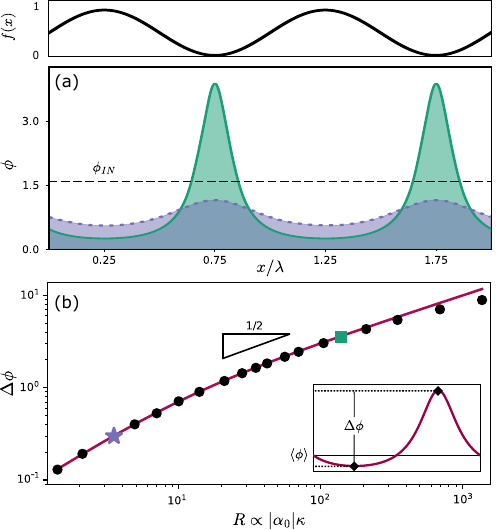}
\caption{\textbf{Density contrast increases with compressibility.}
The black curve above the panels shows the imposed dimensionless activity profile $f(x)$.
\textbf{(a)} Density profiles for low ($\kappa=1.43$, dotted purple) and high ($\kappa=20$, solid green) compressibility. The dashed line marks the isotropic--nematic transition density $\phiIN$.
\textbf{(b)} Density contrast $\Delta\phi$ versus the dimensionless control parameter $R = \alpha_B^0\kappa/(1+\kappa\zeta D)$. Points show numerical solutions of the full two-dimensional model, Eqs.~\eqref{eq:mass}--\eqref{eq:nematic}, and the solid line the analytical solution of the reduced one-dimensional model. The star and square correspond to the profiles in (a). \textit{Inset:} Definition of $\Delta\phi$. Here, \textcolor{red}{$\kappa\in[0.2,200]$}.}
  \label{fig2:contrast}
\end{figure}

\begin{figure}[!t]
  \centering
  \includegraphics[width=\columnwidth]{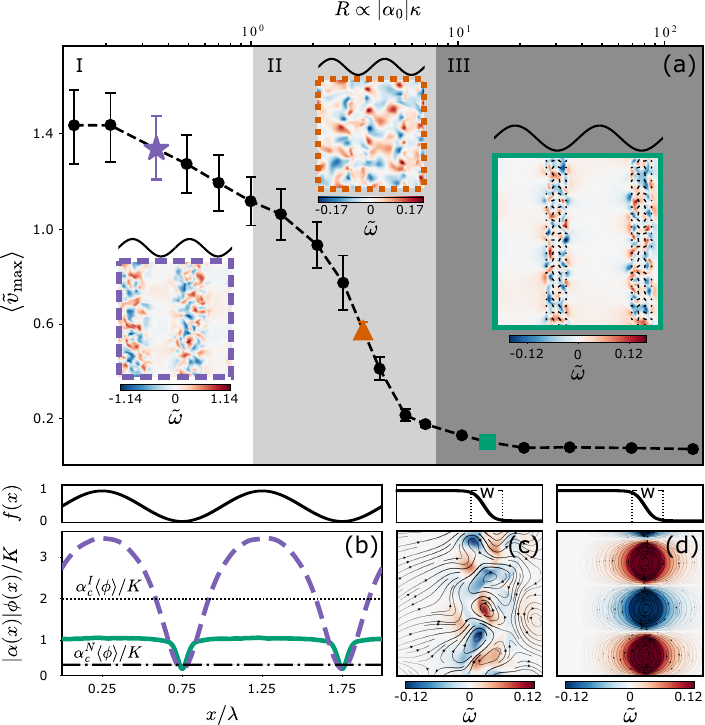}
  \caption{\textbf{Controlling turbulent flows with compressibility.}
\textbf{(a)} Time-averaged maximum speed $\langle\tilde v_{\max}\rangle$
normalized by its incompressible value, versus $R\propto|\alpha_0|\kappa$
for a sinusoidal activity profile with $\lambda=L/2$ and
\textcolor{red}{$\kappa\in[0.2,200]$}.
Regions I--III correspond, respectively, to turbulent flows confined
to high-activity regions, turbulent flows extending throughout the
system, and weak turbulent flows confined to low-activity regions. \textcolor{red}{The boundaries between the shaded regions are set by the
analytical local-threshold predictions for the I--II and II--III crossovers.}
Insets show normalized vorticity $\tilde\omega$ for
\textcolor{red}{$\kappa=0.5$, $5$, and $20$, corresponding to the
purple star, orange triangle, and teal square, respectively}.
For $\kappa=20$, the director field is weighted by the magnitude of
nematic order. \textcolor{red}{See Fig.~S8 for expanded views.}
\textbf{(b)} Effective activity $|\alpha(x)|\phi(x)/K$ for
$\kappa=1.43$ (dashed purple line) and $\kappa=20$ (solid green line).
The dotted and dash-dotted horizontal lines mark
\textcolor{red}{estimates of the thresholds for active turbulence in
the isotropic and nematic states, respectively, evaluated using the
bare Landau coefficient $r_0$}.
\textbf{(c,d)} Normalized vorticity $\tilde\omega$ and streamlines of
the interfacial vortex chain for a hyperbolic-tangent activity profile
of width \textcolor{red}{$W=L/32=16$}, obtained from \textbf{(c)} the numerical solution
and \textbf{(d)} the analytical solution (Sec.~S7).
}
  \label{fig3:dark_vortices}
\end{figure}

We initialize the system in a homogeneous state below the isotropic--nematic transition. Under \emph{uniform} activity, extensile stresses drive the familiar bend instability to active turbulence above a finite, density-dependent threshold, exactly as in incompressible active nematics~\cite{sanchez2012spontaneous,guillamat2016control,decamp2015orientational,doostmohammadi2016stabilization}. Compressibility, however, leaves a clear imprint: the turbulent flow now carries pronounced density inhomogeneities, and because the liquid crystal is lyotropic, the densest regions cross the isotropic--nematic transition and order locally [Fig.~\ref{fig1:result_summary}(a)]. Under \emph{patterned} activity, the isotropic active pressure expels material from high-activity regions and accumulates it in low-activity ones; the resulting contrast is modest at low compressibility [Fig.~\ref{fig1:result_summary}(b)] but pronounced at high compressibility [Fig.~\ref{fig1:result_summary}(c)], thereby relocating the regions where turbulence is localized. In the rest of this Letter we first quantify the patterned-activity density profiles through an exactly solvable 1D reduction (Fig.~\ref{fig2:contrast}), then show how compressibility relocates active turbulence and ultimately traps it into a one-dimensional vortex chain at sharp activity interfaces (Fig.~\ref{fig3:dark_vortices}).

  \vspace{0.1in}
Our work is inspired by experiments on suspensions of active nematogens
immersed in a solvent and confined between two plates a small distance
apart~\cite{Cheng2026,opathalage2019selforganized,chandrakar2020confinement}.
\textcolor{red}{We examine here an effective one-fluid model obtained as the dilute, weak-interphase-coupling
limit of a confined two-fluid suspension (see Sec.~S1 of ~\cite{SupplementalMaterial}). The active nematogens
and passive solvent have distinct in-plane velocities, and
incompressibility constrains only their combined volume flux. Relative
nematogen--solvent motion therefore allows the nematogen phase to compress
while the total suspension remains incompressible.}
We work in terms of the rescaled
\textcolor{red}{nematogen} density $\phi\equiv\rho/\rho_0$
(with $\langle\phi\rangle=1$), the
\textcolor{red}{nematogen} flow velocity $\mathbf{v}$, and the nematic tensor
$Q_{ij}=S(n_in_j-\tfrac12\delta_{ij})$, with $S$ the scalar order parameter
and $\mathbf{n}$ the director. The equations of motion read
\begin{align}
&\partial_t \phi +  \nabla \cdot (\mathbf{v}\, \phi) = D \nabla^2\phi\;,
\label{eq:mass}\\
&\eta \nabla^2\mathbf{v} +\nabla \cdot \boldsymbol{\sigma}^a - \nabla P = \zeta\,\phi\,\mathbf{v}\;,
\label{eq:momentum}\\
&\partial_t \mathbf{Q}  + \textcolor{red}{(\mathbf{v}\cdot\nabla)\mathbf{Q}}= \lambda_N   \mathbf{A} - [\boldsymbol{\omega}, \mathbf{Q}]+ \frac{1}{\gamma}\mathbf{H}\;.
\label{eq:nematic}
\end{align}
The density obeys a continuity equation with diffusivity $D$. The nematic tensor evolves by standard liquid-crystal hydrodynamics, with $A_{ij}=\tfrac12(\partial_iv_j+\partial_jv_i-\delta_{ij}\partial_kv_k)$ the traceless strain rate, $\omega_{ij}=\tfrac12(\partial_iv_j-\partial_jv_i)$ the vorticity, $\lambda_N$ the flow-alignment parameter, and $\gamma$ the rotational viscosity. The flow obeys a screened (Brinkman) Stokes balance, with $\eta$ the shear viscosity and $\zeta$ a friction per unit area  encoding momentum exchange with solvent and substrate, which screens the flow over a length $\ell_s=\sqrt{\eta/\zeta}$. The active stress
\begin{align}
    \Sa = \alpha\,\phi\, \mathbf{Q} + \alpha_B\,\phi\, \mathbb{1}
    \label{eq:sigma-a}
\end{align}
comprises deviatoric and isotropic contributions, both extensile ($\alpha,\alpha_B<0$) and proportional to  $\phi$.  
The \textcolor{red}{nematogen osmotic} pressure is taken as that of an
ideal-gas, $P=\phi/\kappa$, with $\kappa$ the
compressibility\textcolor{red}{, so that $1/\kappa$ is the corresponding
bulk modulus}---the key control parameter below.
\textcolor{red}{In the underlying two-fluid description,
$\boldsymbol{\sigma}^a$ is the stress carried by the nematogen
network, with momentum transferred to the solvent through interphase drag.
The nematogen osmotic pressure $P$ is distinct from the mixture pressure
that enforces incompressibility of the total suspension; accordingly, the
isotropic active stress contributes to the effective pressure of the nematogen phase.} Finally, $\mathbf{H}=-\delta F_{\mathrm{LdG}}/\delta\mathbf{Q}$ derives from a lyotropic Landau--de Gennes free energy $F_{\mathrm{LdG}}=\int d\mathbf{r}\,f_{\mathrm{LdG}}$, with $f_{\mathrm{LdG}}=\tfrac{r(\phi)}{2}\Tr[\mathbf{Q}^2]
+\tfrac{u(\phi)}{4}\Tr[\mathbf{Q}^2]^2
+\tfrac{K}{2}(\nabla\mathbf{Q})^2$ and density-dependent coefficients
$r(\phi)=r_0(1-\phi/\phiIN)$, $u(\phi)=u_0\,\phi/\phiIN$, where
$\phiIN\equiv\rhoIN/\rho_0$ is the isotropic--nematic transition density
and $r_0,u_0,K>0$\textcolor{red}{; for simplicity, we neglect any dependence of $K$ on $\phi$}. Sufficiently dense regions ($\phi>\phiIN$) thus order nematically; we work below the transition on average, $\langle\phi\rangle=1<\phiIN$.

We numerically integrate Eqs.~(\ref{eq:mass}--\ref{eq:nematic}) using a pseudospectral solver on a two-dimensional grid of size $L\times L$~\cite{caballero2024cupss}. Lengths are measured in units of the bare liquid-crystal coherence length $\ell_c=\sqrt{K/r_0}$, times in units of the nematic relaxation time $\tau=\gamma/r_0$, and energies in units of the elastic constant $K$.  All numerics use $\Delta x=\Delta y=1$ and $\Delta t=2\times10^{-4}$ and are run for at least $5\times10^5$ steps, after which the system has achieved steady state. Unless specified otherwise,  parameters are fixed to $\alpha=-5.0$, $\alpha_B=-3.5$, $D=5.0$, $u_0=5.0$, $\langle\phi\rangle=1$, $\phi_{IN}=1.43$, $\eta=1.0$, $\lambda_N=1$, $\zeta=10^{-4}$, and $L=512$. See SI Sec.~S6 for additional numerical details. All reported steady-state observables are time-averaged.

  \vspace{0.1in}
For uniform activity, our compressible extensile active liquid crystal is linearly unstable to bend deformations driven by the active deviatoric stress, just like its incompressible counterpart. An important difference is that in the compressible fluid turbulent flows are accompanied by large density fluctuations. The isotropic state of the active liquid crystal is linearly unstable when the magnitude of activity exceeds a critical value, given by 
$\alpha_c^I
= \,\frac{2\zeta K }{\gamma\lambda_N}\;
\left(1 + \ell_s/\xi\right)^{2}$, with $\xi=\sqrt{K/r(\langle\phi\rangle)}$ the liquid crystal correlation length. This threshold remains finite in the thermodynamic limit and depends on the distance of $\langle\phi\rangle$ from $\phiIN$. The longitudinal sector couples density and nematic order and supports propagating waves (see SI Sec.~S2).
In the nematic phase ($\langle\phi\rangle > \phiIN$), instead, the system becomes unstable for  $|\alpha|>\alpha_c^N
=
\frac{2\zeta K}{\gamma(\lambda_N+1)}$ \cite{simha2002hydrodynamic,giomi2015geometry,shankar2019hydrodynamics,alert2022active,srivastava2016negative,thampi2016active, ngo2014large}.

\textcolor{red}{
Density correlations and number fluctuations under uniform illumination are
analyzed in Sec.~S4. In the homogeneous isotropic state, fluctuations are
normal, with $S_\phi(q\to0)\sim q^0$ and $\Delta N\sim N^{1/2}$
~\cite{simha2003active,mishra2006active,marchetti2013hydrodynamics,baskaran2008hydrodynamics}. In isotropic \emph{compressible} fluids  dense regions can acquire nematic order at sufficiently large activity and compressibility.
 In the homogeneous nematic state, the screening length
$\ell_s=\sqrt{\eta/\zeta}$ separates two regimes: for
$q\ll\ell_s^{-1}$, coupling of the soft director mode to density recovers
the giant-number-fluctuation behavior of dry active nematics
~\cite{simha2003active,mishra2006active,narayan2007long,chate2006simple,ngo2014large,kozhukhov2024mitigating,zhang2010collective},
with $S_\phi(q)\sim q^{-2}$ and $\Delta N\sim N$; for
$q\gg\ell_s^{-1}$, viscous dissipation restores
$S_\phi(q)\sim q^0$ and $\Delta N\sim N^{1/2}$.
}



\textcolor{red}{
We consider spatial patterns of activity,
$\alpha(x)=-\alpha_0 f(x)$ and
$\alpha_B(x)=-\alpha_B^0 f(x)$, with
$\alpha_B^0=0.7\,\alpha_0$ and $\alpha_0>0$.
Because $\alpha$ and $\alpha_B$ are the deviatoric and isotropic components
of the same underlying active stress, taking their amplitudes to be
proportional provides a simple default parametrization.
An exactly solvable one-dimensional reduction with $\mathbf{Q}=0$
captures the resulting density redistribution. Combining continuity with
overdamped Stokes balance, Eqs.~\eqref{eq:mass} and~\eqref{eq:momentum},
yields an advection--diffusion equation
}
\begin{equation}
    \partial_t \phi
    =
    \partial_x
    \left[
        -v_d(x)\,\phi
        +
        D_e(x)\,\partial_x\phi
    \right]\;,
    \label{eq:adv-diff}
\end{equation}
with drift $v_d(x)=\partial_x\alpha_B(x)/\zeta$ and effective diffusivity $D_e(x)=D_p+D_a(x)$ built from competing passive and active contributions,
    $D_p=D+(\kappa\zeta)^{-1}$
   and
    $D_a(x)=-\alpha_B(x)/\zeta$.
The steady-state flux vanishes and $\phi(x)\propto 1/D_e(x)$: material accumulates wherever it moves slowly. 
The resulting advection--diffusion equation is closely related to descriptions of bacterial transport in spatially heterogeneous environments, where position-dependent motility or reorientation rates generate nonuniform steady-state density profiles~\cite{schnitzer1993theory,solon2018generalized,fily2012athermal,marchetti2016minimal}. Related density patterning has also been realized experimentally in light-controlled \textit{E.~coli} suspensions~\cite{frangipane2018dynamic}.

The steady state reads
\begin{align}
   \phi(x)
   =
   \frac{C}{1+R\,f(x)}\;,
   \qquad
   C
   =
   \left\langle
   \frac{1}{1+R\,f}
   \right\rangle^{-1},
    \label{eq:rho-profile}
\end{align}
where $C$ is fixed by number conservation, $\langle\phi\rangle=L^{-1}\int_0^L dx\,\phi(x)=1$.  The profile is governed by the single dimensionless ratio $R=D_a^0/D_p=\alpha_B^0\kappa/(1+\kappa\zeta D)\simeq\alpha_B^0\kappa$, where $D_a^0=\alpha_B^0/\zeta$ and the last approximation holds when $\kappa\zeta D\ll1$.
For a sinusoidal profile $f(x)=[1+\sin(2\pi x/\lambda)]/2$, we find $C=\sqrt{1+R}$. 

 \textcolor{red}{
The dimensionless number $R$ has two equivalent interpretations: it is an
activity P\'eclet number comparing active transport to passive relaxation,
and, writing $B=1/\kappa$, $1+R\simeq(B+\alpha_B^0)/B$ measures the active
pressure relative to the passive bulk modulus.
}

For sinusoidal activity, the density contrast is given by $\Delta\phi=\phi_{\max}-\phi_{\min}=R/\sqrt{1+R}$, independent of the modulation wavelength (see Sec.~S5B). It grows linearly, $\Delta\phi\simeq R$ for $R\ll1$, and as $\Delta\phi\sim\sqrt{R}$ for $R\gg1$, where the high-activity regions fully evacuate ($\phi_{\min}\to0$). The 1D prediction matches the density contrast of the full model (Eqs.~(\ref{eq:mass}--\ref{eq:nematic})) for a given value of $\lambda$ with no fitting (Fig.~\ref{fig2:contrast}(b)), establishing compressibility as the key control parameter for density variations.

This agreement, however, fails when the activity is modulated on wavelengths smaller than the typical vorticity correlation length 
$\ell_\omega$. When $\lambda<\ell_\omega$ the deviatoric stress and nematic elasticity become comparable to the isotropic active pressure. Vortical flows generated by deviatoric active stresses then continuously mix up the density. This
is evident in Fig.~S5 that shows the density contrast $\Delta\phi$ obtained numerically as a function of $\lambda$: $\Delta\phi$ approaches the analytical result from the scalar model at large $\lambda$, but it is strongly suppressed at small wavelengths. Throughout this work we use $\lambda=L/2$, which lies in the large-wavelength regime where the scalar model prediction captures the density contrast.

\vspace{0.1in}

\textcolor{red}{
By tuning compressibility at fixed activity magnitude, turbulent flow shifts
from high- to low-activity regions while its overall magnitude decreases
[Fig.~\ref{fig3:dark_vortices}(a)]. This behavior follows from the local
effective active strength $|\tilde{\alpha}(x)|=|\alpha(x)|\phi(x)$
[Fig.~\ref{fig3:dark_vortices}(b)]. At low compressibility,
$|\tilde{\alpha}(x)|$ exceeds the isotropic threshold $\alpha_c^I$ only in
the high-activity regions. At high compressibility, density accumulation
drives the low-activity regions into the nematic phase, where the lower
threshold $\alpha_c^N$ is exceeded, while depletion drives the high-activity
regions below $\alpha_c^I$. Turbulence therefore relocates entirely to the
low-activity regions.
}

Sharp gradients of activity provide a mechanism for an intriguing dynamical state: a chain of counter-rotating  vortices localized at the high/low-activity interface. At high compressibility and for a range of activity, the effective activity $|\tilde{\alpha}|$ can remain below the threshold for active turbulence in both the low- and high- activity regions and vortical flows can only appear at the  nematic high/low-activity interface. 

More precisely, note that vortical flows in Eqs.~\eqref{eq:mass}--\eqref{eq:nematic} are solely sourced by the deviatoric active forcing $\sim \mathbf{\nabla}\cdot(\alpha \phi \mathbf{Q})$ (Sec.~S7). Nematic order $\mathbf{Q}$ is finite and roughly uniform throughout the low-activity region, whereas $\alpha \phi$ is largest at its edge (Fig.~\ref{fig3:dark_vortices}(b)). As a result, the dominant active forcing, $\sim \mathbf{Q}\cdot \nabla(\alpha \phi)$, becomes localized at the interface, where the gradient of $\alpha \phi$ is largest. In this sense, activity acts as a form of ``soft confinement'': the \textit{narrower the interfacial region}, the more localized the vortical flow, until a one-vortex-thick chain is formed.

\nocite{Shankar2024,mozaffari2021defect}
The simplest way to obtain the one-dimensional vortex-chain state is to use activity profiles with sharp gradients, such as a square pulse regularized by hyperbolic tangents. In this case, the activity vanishes exactly in the low-activity region. For sufficiently large compressibility, when the effective activity strength in the high-activity region remains below $\alpha_c^I$ (Fig.~\ref{fig3:dark_vortices}(b)), vortices cannot be sustained in the bulk of either region and are therefore confined to the nematic high/low-activity interface (Fig.~\ref{fig3:dark_vortices}(c)) \footnote{The incompressible frameworks of Refs.~\cite{Shankar2024,mozaffari2021defect} suggest that related vortex or defect confinement could, in principle, be used to generate vortex chains when the width of the active region is comparable to the intrinsic vortex correlation length. Here, by contrast, the chain forms for any activity-pattern wavelength larger than approximately twice the vortex correlation length.}.
Although one-dimensional vortex chains have been discussed previously in systems confined by rigid walls~\cite{voituriez2005spontaneous,shendruk2017dancing,gulati2022boundaries}\textcolor{red}{~\cite{degraafsousa2026spontaneous}}, the state described here arises instead in a free-standing fluid, where soft confinement is purely imposed by the activity pattern. The high/low-activity interface acts as a soft and reconfigurable boundary that localizes the flow without the need for solid walls. This makes the resulting vortex chain both controllable and mobile in principle, opening the possibility of transporting suspended objects.

\textcolor{red}{
To further quantify this state, we analyze a high/low-activity interface of
width $W$, parametrized by
$f(x)=\tfrac{1}{2}[1+\tanh(x/W)]$ (Fig.~S10).
Adapting the stability analysis of Ref.~\cite{valani2026competing}
(see SI, Sec.~S7), we identify a confinement-controlled finite-wavelength
mechanism with $q_c=O(W^{-1})$. Assuming a sinusoidal director modulation along the
interface and the density profile from our 1D model, we then solve the
Stokes equation, Eq.~\eqref{eq:momentum}, approximating $\phi\simeq1$ in
the frictional drag. The resulting vorticity has the form
$\omega(x,y)=A(q,x)\sin(qy)$, with
$A(q,x)\sim e^{-\mu_q|x|}$ and
$\mu_q=\sqrt{q^2+\ell_s^{-2}}$.
Thus the stability analysis identifies the confinement-controlled scaling
$q_c=O(W^{-1})$, while the post-instability solution determines the transverse
localization:
$\mu_{q_c}^{-1}=(q_c^2+\ell_s^{-2})^{-1/2}\sim W$ for $\ell_s\gg W$.
The resulting vortex chain is observed directly in the numerical solution
[Fig.~\ref{fig3:dark_vortices}(c)].
}

In conclusion, we have developed an effective one-fluid model of a  compressible active liquid crystal that accounts for the large density variations observed in active liquid crystal suspensions, such as the light-responsive microtubule--kinesin system of Refs.~\cite{Cheng2026,opathalage2019selforganized,chandrakar2020confinement}. For uniform activity, the dynamics is controlled by the familiar extensile bend instability that drives active turbulence in incompressible fluids~\cite{simha2002hydrodynamic,marenduzzo2007steady,edwards2009spontaneous,giomi2015geometry,doostmohammadi2018active,alert2022active}.
For spatially varying activity, compressibility and isotropic active stresses  control the magnitude of the resulting density variations. Remarkably, a scalar version of the hydrodynamic model captures both the density profiles and the large density contrast obtained numerically. By using compressibility as a control parameter, we uncover a mechanism for controlling and localizing active turbulent flows and show that sharp interfaces can stabilize  a one-dimensional chain of counter-rotating vortices.
Unlike previously discussed vortex chains and dynamically ordered vortex lattices generated by hard-wall channel confinement~\cite{voituriez2005spontaneous,shendruk2017dancing,gulati2022boundaries}, the one-dimensional vortex chain found here emerges in a free-standing fluid through reconfigurable, activity-induced soft confinement, thereby providing a potential route toward controllable transport. Our analytical results and numerical solutions therefore
\textcolor{red}{provide a framework for understanding}
the experimental observations in patterned active nematic suspensions~\cite{Cheng2026}
\textcolor{red}{and for}
controlling their density organization and flow.

\begin{acknowledgments}
We thank Fernando Caballero, Yu-Chuan Cheng, Zvonimir Dogic, Michael Hagan, and Sriram Ramaswamy for insightful discussions. M.C.M. and D.K. acknowledge support from the National Science Foundation through DMREF Award No.~2324194. The derivation of the effective one-fluid model of a compressible active nematic (M.C.M.) was supported by NSF DMR-2528734. D.K. was additionally supported by the Dutch Research Council (NWO) through Rubicon Grant No.~019.251EN.011 (Grant ID: 10.61686/LAKBE17140). A.B. was supported in part by NSF Grant No.~PHY-2309135 through the Kavli Institute for Theoretical Physics (KITP), NSF Grant No.~DMR-2202353, and NSF Grant No.~2011846.  P.G. was supported, in part, by the Tarbutton Interdisciplinary Postdoctoral Fellowship at Emory College of Arts and Sciences.
\end{acknowledgments}
\nocite{ChapmanCowling1970,LopezDeHaro1983,Garzo2007,vafa2021fluctuations,krommydas2026compressible}

\bibliography{bib}
\end{document}


\title{Supplemental Material for: Controlling Turbulent Flows in Compressible Active \textcolor{red}{Liquid Crystals}}

\author{Dimitrios Krommydas}
\affiliation{Department of Physics, University of California Santa Barbara, CA 93106, USA}

\author{Paarth Gulati}
\affiliation{Department of Physics, Emory University, Atlanta, GA, USA}
\affiliation{Initiative in Theory and Modeling of Living Systems, Emory University, Atlanta, GA, USA}

\author{Aparna Baskaran}
\affiliation{Department of Physics, Brandeis University, 415 South Street, Waltham, Massachusetts 02453, USA}

\author{M. Cristina Marchetti}
\affiliation{Department of Physics, University of California Santa Barbara, CA 93106, USA}
\affiliation{Interdisciplinary Program in Quantitative Biosciences, University of California Santa Barbara, Santa Barbara, CA 93106, USA}

\date{\today}

\maketitle

\section{Model}

\REVSTART

We derive our effective compressible active liquid crystal model as a
systematic reduction of a two-fluid description of a quasi-2D active
suspension. The system consists of two mechanically distinct components:
active nematogens and a passive solvent, confined between two plates
separated by a small distance $h$. We choose a coordinate system with the $z$ axis normal to the planes.

Let $\Phi_n$ and $\Phi_s=1-\Phi_n$ denote the local volume fractions of
nematogens and solvent, respectively. We denote the in-plane velocity of the
nematogens by $\mathbf v_n$ and decompose the solvent velocity as
$\mathbf v_s=(\mathbf v_{s}^{\perp},v_{sz})$. We neglect nematogen flow in the $z$ direction. Throughout this section,
$\nabla=(\partial_x,\partial_y)$ and
$\nabla^2=\partial_x^2+\partial_y^2$ denote in-plane derivatives, while the
full three-dimensional Laplacian is
$\nabla^2+\partial_z^2$. Each constituent is taken to be materially
incompressible, but the two phases are allowed to move relative to one
another. Their conservation equations are therefore
\begin{align}
\partial_t\Phi_n+\nabla\cdot(\Phi_n\mathbf v_n)&=0,
\label{eq:twofluid_mass_n}\\
\partial_t\Phi_s+\nabla\cdot(\Phi_s\mathbf v_{s}^{\perp})
+\partial_z(\Phi_s v_{sz})&=0.
\label{eq:twofluid_mass_s}
\end{align}
We define the gap-averaged in-plane solvent velocity as
\begin{equation}
\overline{\mathbf v}_s
=
\frac{1}{h}\int_0^h dz\,\mathbf v_{s}^{\perp}.
\end{equation}
Since the confining plates are impermeable,
$v_{sz}|_{z=0}=v_{sz}|_{z=h}=0$. Upon integrating
Eq.~\eqref{eq:twofluid_mass_s} across the gap, the out-of-plane flux
therefore vanishes as a boundary term. Adding the resulting gap-averaged
solvent equation to Eq.~\eqref{eq:twofluid_mass_n} gives the
incompressibility constraint on the combined in-plane volume flux,
\begin{equation}
\nabla\cdot\left(
\Phi_n\mathbf v_n+\Phi_s\overline{\mathbf v}_s
\right)=0.
\label{eq:twofluid_incompressibility}
\end{equation}
Thus, incompressibility of the total suspension does not imply
$\nabla\cdot\mathbf v_n=0$. Local compression of the nematogen phase can be
accommodated by a compensating solvent counterflow. This relative motion is
the compressible degree of freedom retained in the effective model below.

At low Reynolds number, the two phases satisfy separate force balance equations. We
assume that the nematogen fields are approximately uniform across the narrow
gap, so that $\partial_z\Phi_n=0$ and the nematogen velocity and stress have
no leading-order $z$ dependence, while retaining their full spatial
dependence in the $(x,y)$ plane. The in-plane force balances are then
\begin{align}
\nabla\cdot\bm{\sigma}_n
-\Phi_n\nabla p
-\xi_d(\mathbf v_n-\mathbf v_{s}^{\perp})
-\zeta_w\Phi_n\mathbf v_n
&=0,
\label{eq:twofluid_force_n}\\
\eta_s\left(\nabla^2+\partial_z^2\right)\mathbf v_{s}^{\perp}
-\Phi_s\nabla p
+\xi_d(\mathbf v_n-\mathbf v_{s}^{\perp})
&=0.
\label{eq:twofluid_force_s}
\end{align}
where $p$ is the common pressure enforcing
Eq.~\eqref{eq:twofluid_incompressibility}, $\eta_s$ is the solvent
viscosity, $\xi_d$ is the interphase friction coefficient, and $\zeta_w$
denotes a possible direct friction of the nematogens with the confining
boundaries. 
Although the solvent can have a nonzero normal velocity $v_{sz}$ within the
gap, this component is smaller than the in-plane solvent velocity by the
thin-gap aspect ratio. The $z$ component of the solvent force balance
therefore gives $\partial_z p=0$ to leading order in the thin-gap expansion,
so that the pressure appearing above depends only on the in-plane
coordinates.
We take the minimal symmetric form of the interphase friction
\begin{equation}
\xi_d=\xi_d^0\Phi_n\Phi_s,
\label{eq:interphase_drag}
\end{equation}
which reduces to $\xi_d\simeq\xi_d^0\Phi_n$ in the dilute limit ($\Phi_s\simeq 1$). The
nematogen stress is written as
\begin{equation}
\bm{\sigma}_n
=
\Sv
+\Phi_n\left(a\mathbf Q+a_B\mathbb{1}\right)
-\Pi(\Phi_n)\mathbb{1},
\label{eq:nematogen_stress}
\end{equation}
where $\Pi(\Phi_n)$ is the osmotic pressure of the nematogen phase and
$a$ and $a_B$ are the deviatoric and isotropic active stresses per unit
nematogen volume fraction.
We assign the motor-generated active contribution to the nematogen-phase
stress because, over the regime described by the continuum theory, the
filaments form an interacting active network. Motor-generated force dipoles
therefore act within this network, while momentum is transferred to the
surrounding solvent through interphase drag. Here the dilute limit refers to
a small nematogen volume fraction within this network regime. At very low
concentrations, where active filaments or bundles become isolated, a
different coarse-grained description in which the active forcing acts
directly on the surrounding fluid would be more appropriate.

We now exploit the quasi-2D geometry. The nematogens move only in plane,
while the solvent obeys no-slip boundary conditions at the two plates. If
$\ell$ denotes a characteristic in-plane length scale, then
$\nabla^2\mathbf v_{s}^{\perp}\sim\mathbf v_{s}^{\perp}/\ell^2$, whereas
$\partial_z^2\mathbf v_{s}^{\perp}\sim\mathbf v_{s}^{\perp}/h^2$. Thus, in the
thin-gap limit ($h\ll\ell$) the in-plane viscous term is smaller by a factor
of order $(h/\ell)^2$ and can be neglected to leading order. The in-plane
solvent force balance therefore becomes
\begin{equation}
\eta_s\partial_z^2\mathbf v_{s}^{\perp}
-\xi_d\mathbf v_{s}^{\perp}
=
\Phi_s\nabla p-\xi_d\mathbf v_n.
\label{eq:solvent_gap}
\end{equation}
Solving Eq.~\eqref{eq:solvent_gap} with
$\mathbf v_{s}^{\perp}|_{z=0}
=\mathbf v_{s}^{\perp}|_{z=h}=0$ and averaging across the gap gives a
Darcy--Brinkman relation for the gap-averaged solvent velocity
$\overline{\mathbf v}_s$,
\begin{equation}
\Gamma\overline{\mathbf v}_s
=
\xi_d\left(\mathbf v_n-\overline{\mathbf v}_s\right)
-\Phi_s\nabla p,
\label{eq:darcy_brinkman}
\end{equation}
where
\begin{equation}
\Gamma
=
\xi_d\,\frac{1-g(h/2\ell_v)}{g(h/2\ell_v)},
\qquad
g(x)=1-\frac{\tanh x}{x},
\qquad
\ell_v=\sqrt{\frac{\eta_s}{\xi_d}}.
\label{eq:Gamma}
\end{equation}
For weak interphase coupling, $h\ll\ell_v$, this reduces to the usual
Hele--Shaw friction
\begin{equation}
\Gamma\simeq\Gamma_0=\frac{12\eta_s}{h^2}.
\label{eq:Gamma0}
\end{equation}

The gap-averaged solvent velocity can be eliminated exactly from
Eq.~\eqref{eq:darcy_brinkman},
\begin{equation}
\overline{\mathbf v}_s
=
\frac{\xi_d\mathbf v_n-\Phi_s\nabla p}{\Gamma+\xi_d}.
\label{eq:solvent_eliminated}
\end{equation}
Substituting this result into the nematogen force balance and using the
gap-averaged form of Eq.~\eqref{eq:twofluid_incompressibility} gives
\begin{align}
\nabla\cdot\bm{\sigma}_n
-\zeta_{\mathrm{eff}}\mathbf v_n
-\chi\nabla p&=0,
\label{eq:twofluid_effective_force}\\
\nabla\cdot\left(\mathcal K\nabla p\right)
&=
\nabla\cdot\left(\chi\mathbf v_n\right),
\label{eq:twofluid_pressure}
\end{align}
with
\begin{equation}
\zeta_{\mathrm{eff}}
=
\zeta_w\Phi_n+\frac{\xi_d\Gamma}{\Gamma+\xi_d},
\qquad
\chi
=
\Phi_n+\frac{\xi_d\Phi_s}{\Gamma+\xi_d},
\qquad
\mathcal K
=
\frac{\Phi_s^2}{\Gamma+\xi_d}.
\label{eq:twofluid_coefficients}
\end{equation}
Equations~\eqref{eq:twofluid_effective_force} and
\eqref{eq:twofluid_pressure} make explicit the two effects generated by the
passive solvent: a density-dependent friction and a pressure-mediated
feedback associated with the solvent counterflow required by total-volume
conservation.

The local equation used in the main text follows as the leading dilute,
weak-interphase-coupling limit of these equations, i.e.,
$\Phi_n\ll1$, $\Phi_s\simeq 1$, $\xi_d/\Gamma_0\ll1$, and
$\zeta_{\mathrm{eff}}\simeq(\xi_d^0+\zeta_w)\Phi_n$, while the additional
solvent-pressure feedback is subleading. The leading nematogen force balance
then becomes
\begin{equation}
\nabla\cdot\left[
\Sv
+\Phi_n(a\mathbf Q+a_B\mathbb{1})
-\Pi(\Phi_n)\mathbb{1}
\right]
=
(\xi_d^0+\zeta_w)\Phi_n\mathbf v_n.
\label{eq:twofluid_weak_limit}
\end{equation}
Defining the normalized concentration
$\phi=\Phi_n/\overline{\Phi}_n=\rho/\rho_0$, where $\rho$ is the
nematogen partial density and $\rho_0$ its mean value,  writing
$\mathbf v\equiv\mathbf v_n$ for the nematogen velocity, together with
\begin{equation}
\alpha=\overline{\Phi}_n a,
\qquad
\alpha_B=\overline{\Phi}_n a_B,
\qquad
\zeta=\overline{\Phi}_n(\xi_d^0+\zeta_w),
\end{equation}
and writing the local nematogen osmotic pressure as
$P(\phi)=\Pi(\overline{\Phi}_n\phi)$, we obtain the effective equations used
throughout the main text. Separating an explicit diffusive contribution from
the coarse-grained osmotic response gives
\begin{equation}
\partial_t \phi + \nabla \cdot (\phi\mathbf{v}) = D\nabla^2\phi.
\label{eq:mass}
\end{equation}
In the overdamped scalar sector, this explicit diffusion combines with the
osmotic-pressure contribution into an effective diffusivity.

The equilibrium physics is governed by a density contribution together with
a lyotropic liquid crystal free energy,
\begin{equation}
F = \int d\mathbf{r}\left[\frac{\phi}{\kappa}\log\phi
+ \frac{r_0}{2}\left(1 - \frac{\phi}{\phiIN}\right)\mathrm{Tr}[\mathbf{Q}^2]
+ \frac{u_0}{4}\left(\frac{\phi}{\phiIN}\right)\mathrm{Tr}[\mathbf{Q}^2]^2
+ \frac{K}{2}(\nabla\mathbf{Q})^2\right]\;,
\label{eq:free energy}
\end{equation}
where $\phiIN$ is the concentration of the isotropic--nematic transition.
The first term gives the ideal osmotic pressure $P=\phi/\kappa$.

Taking the nematogen viscous stress to have constant shear viscosity
$\eta$ and zero bulk viscosity, so that
$\nabla\cdot\Sv=\eta\nabla^2\mathbf v$, the in-plane nematogen flow is
governed by the Stokes equation,
\begin{equation}
\eta\nabla^2\mathbf{v}
+\nabla\cdot\Sa
-\nabla P
=
\zeta\phi\mathbf{v},
\label{eq:stokes}
\end{equation}
with
\begin{equation}
\Sa
=
\alpha\phi\mathbf Q+\alpha_B\phi\mathbb{1}.
\label{eq:active_stress}
\end{equation}
Here $\mathbf v$ is the velocity of the active nematogen phase. The
coefficient $\zeta$ contains the leading interphase drag and any direct wall
contribution, while $P$ is the constitutive osmotic pressure of the
nematogen phase. It should therefore be distinguished from the mixture
pressure $p$ appearing in
Eqs.~\eqref{eq:twofluid_force_n}--\eqref{eq:twofluid_pressure}, which acts as
the Lagrange multiplier enforcing incompressibility of the total suspension
and has been eliminated in obtaining Eq.~\eqref{eq:stokes}.

The isotropic active stress contributes the gradient
$\nabla\cdot(\alpha_B\phi\mathbb{1})=\nabla(\alpha_B\phi)$, with any imposed
spatial dependence of the activity understood. It is therefore an extensile active pressure and can be combined
algebraically with the nematogen osmotic pressure,
$\nabla\cdot\left(\alpha_B\phi\mathbb{1}\right)-\nabla P
=
-\nabla\left[P-\alpha_B\phi\right].
$
In an incompressible one-fluid theory, a gradient of this form can be
absorbed into the Lagrange-multiplier pressure. The situation is different
here because the active stress is part of the nematogen-phase stress rather
than a stress of the total mixture. If $\alpha_B\phi\mathbb{1}$ acted only
as a stress of the total incompressible suspension, its gradient could
indeed be absorbed into the mixture pressure $p$ and would not drive
relative motion. Instead, it contributes to the force balance of the
nematogen phase and can be viewed as an active contribution to its
constitutive pressure, $P_{\rm eff}=P-\alpha_B\phi$. Since the nematogen
velocity is not constrained to be divergence-free, gradients of
$P_{\rm eff}$ drive relative nematogen--solvent motion and therefore
concentration redistribution.

Finally, the nematic order parameter $\mathbf Q$, carried solely by the
active phase, evolves according to standard liquid crystal hydrodynamics,
\begin{equation}
\partial_t\mathbf{Q}
+
(\mathbf{v}\cdot\nabla)\mathbf{Q}
=
\lambda_{N}\mathbf{A}
-[\bm{\omega},\mathbf{Q}]
+\frac{1}{\gamma}\mathbf{H},
\label{eq:nematic}
\end{equation}
where $\mathbf{A}_{ij} =
\frac{1}{2}(\partial_i v_j+\partial_j v_i-\delta_{ij}\partial_kv_k)$ is the
strain rate tensor, $\bm{\omega}_{ij} =
\frac{1}{2}(\partial_i v_j-\partial_j v_i)$ is the vorticity tensor,
$\mathbf{H}=-\delta F/\delta\mathbf{Q}$ is the molecular field,
$\lambda_N$ is the flow-alignment parameter, and $\gamma$ is the rotational
viscosity. Since $\mathbf Q$ is an intensive order parameter of the
nematogen phase, it is advected with the material derivative associated with
$\mathbf v$. We use gradients of the nematogen velocity in
$\mathbf A$ and $\bm{\omega}$ as a constitutive choice of the minimal
active-liquid-crystal model. Passive elastic stresses are neglected since
the hydrodynamics is dominated by the active and dissipative terms.

For geometries that reduce by symmetry to a single spatial coordinate,
including one-dimensional stripes and axisymmetric circular patterns,
Eq.~\eqref{eq:twofluid_pressure} can be integrated locally even when
$\xi_d/\Gamma_0$ is not small. For the one-dimensional case considered
here, zero total through-flow gives
\begin{equation}
\mathcal K\,\partial_xp=\chi v_x,
\end{equation}
and therefore an effective local friction
\begin{equation}
\zeta_{\mathrm{1D}}
=
\zeta_{\mathrm{eff}}+\frac{\chi^2}{\mathcal K}
=
(\xi_d^0+\zeta_w)\Phi_n+\mathcal O(\Phi_n^2)
\end{equation}
in the dilute limit. Thus the density profile obtained with the simplified scalar model, which neglects the liquid-crystalline degrees of freedom, does not
require weak interphase coupling, whereas the local form of the full
two-dimensional model follows from the additional weak-coupling
approximation.

Together, Eqs.~\eqref{eq:mass}, \eqref{eq:stokes}, and
\eqref{eq:nematic} constitute our effective minimal model of a compressible
active liquid crystal.

\REVEND

\section{Linear Stability of the Homogeneous, Isotropic State}

Next, we examine the linear stability of an isotropic quiescent state with $\phi=\langle\phi\rangle$, $\mathbf{v}=0$ and $\mathbf{Q}=0$.  
The equations for linear fluctuations $\delta\phi=\phi-\langle\phi\rangle$, $\delta Q_{ij}=Q_{ij}$ and $\delta v_i=v_i$ are given by
\begin{equation}
\begin{split}
    &\partial_t  \delta\phi + \partial_i  v_i =  D\nabla^2  \delta\phi\;, \\
    &\zeta v_i = \eta \nabla^2  v_i - \frac{1}{\kappa} \partial_i  \delta\phi + \alpha \partial_j  Q_{ij} + \alpha_B \partial_i  \delta\phi \;,\\
    &\partial_t  Q_{ij} = \frac{\lambda_N }{2} \left(\partial_i  v_j + \partial_j  v_i - \partial_k  v_k \delta_{ij}\right) - \frac{\tilde{r}}{\gamma} Q_{ij} + \frac{K}{\gamma}\nabla^2  Q_{ij}\;.
\label{eq:model_linear}
\end{split}
\end{equation}
where $\tilde{r}\equiv r(\langle\phi\rangle)=r_0(1-\langle\phi\rangle/\phiIN)$ is the Landau coefficient evaluated at the homogeneous state. Note that, unlike for incompressible flows, the strain-rate tensor in the flow alignment term needs to be explicitly made traceless by subtraction of its trace.

We define the Fourier transform by
\begin{equation}
A(\mathbf{r},t)=\int_{\mathbf q} e^{i\mathbf q\cdot\mathbf r} A(\mathbf q,t),
\qquad
\int_{\mathbf q}\equiv \int \frac{d^2q}{(2\pi)^2}.
\end{equation}
Given the isotropy of the reference state, components longitudinal and transverse to $\mathbf q$ decouple. We define
\begin{equation}
v_\parallel=\hat q_i v_i,
\qquad
v_\perp=\varepsilon_{ij}\hat q_i v_j,
\qquad
Q_{\parallel}\equiv \hat{q}_i\hat{q}_j  Q_{ij},
\qquad
Q_{\perp} \equiv \varepsilon_{is}\,\hat q_s \,\hat q_j \, Q_{ij},
\end{equation}
where $\varepsilon_{ij}$ is the Levi-Civita tensor.

After Fourier transforming and solving for the velocity, we obtain
\begin{equation}
   i q_i  v_j =
   \frac{1}{\zeta+ \eta q^2}
   \left(
   -\alpha \left[ Q_{\parallel} q_i q_j
   +
   Q_{\perp} q_i \varepsilon_{j\ell} q_\ell \right]
   -
   \left(\alpha_B - \frac{1}{\kappa}\right) q_i q_j \delta\phi
   \right)\;.
   \label{eq:vq}
\end{equation}
Using Eq.~\eqref{eq:vq} to eliminate the velocity, the equations for density and Q-tensor fluctuations are given by
\begin{align}
\partial_t \delta\phi
&= \left[-\,D \,q^2
+ \left(\alpha_B - \frac{1}{\kappa}\right)
\frac{q^2}{\zeta+\eta q^2}\right]\delta\phi
+ \alpha\,\frac{q^2}{\zeta+\eta q^2}\,Q_{\parallel}\;,
\label{eq:phi}
\\
\partial_t Q_{\parallel}
&= -\left(\alpha_B - \frac{1}{\kappa}\right)
\frac{\lambda_N  q^2}{2\bigl(\zeta+\eta q^2\bigr)}
\,\delta\phi
- \left(
\frac{\tilde r}{\gamma}
+
\alpha\,\frac{\lambda_N q^2}{2\bigl(\zeta+\eta q^2\bigr)}
+
\frac{K}{\gamma}q^2
\right)Q_{\parallel}\;,
\label{eq:Qparallel}\\
&\partial_t Q_{\perp}
= -\left[
\frac{\tilde r}{\gamma}
+
q^2\left(
\frac{K}{\gamma}
+
\frac{\lambda_N \alpha}{2\bigl(\zeta +\eta q^2\bigr)}
\right)
\right]Q_{\perp}.
\label{eq:stability_perp_SI}
\end{align}

\subsection{Transverse Component}

The transverse component $Q_\perp$ does not couple to  density fluctuations (Eq.~\eqref{eq:stability_perp_SI}) and becomes unstable via the familiar \textit{extensile} active nematic bend instability discussed in previous literature~ \cite{simha2002hydrodynamic,giomi2015geometry,shankar2019hydrodynamics,alert2022active,srivastava2016negative,thampi2016active, ngo2014large,vafa2021fluctuations}
that leads to active turbulence. The growth rate of $Q_\perp$ fluctuation can be read off Eq.~\eqref{eq:stability_perp_SI} and is given by
\begin{equation}
    i \omega_{\perp}(q)
    =
    - \left(
    \frac{\tilde{r}}{\gamma}
    +
    \frac{K q^2}{\gamma}
    +
    \frac{q^2 \lambda_N \alpha}{2( \zeta  + \eta   q^2)}
    \right)\;.
\label{eq:tranverse_disperion}
\end{equation}
With our convention, the system is unstable when $i\omega$ is positive  and is stable when $i\omega$ is negative.

\begin{figure*}[htpb!]
    \centering
    \includegraphics[width=0.9\textwidth]{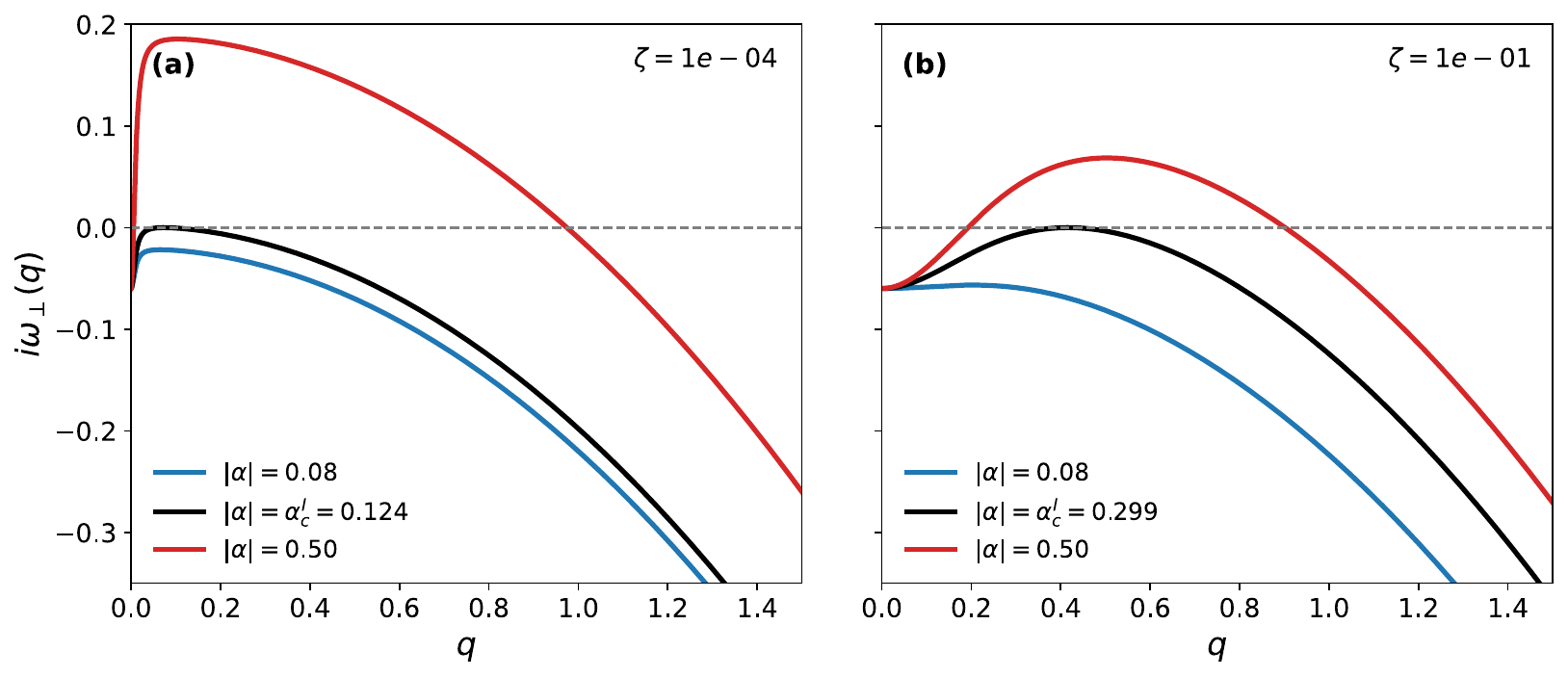}
    \caption{
    Dispersion relation for the growth rate of transverse order-parameter fluctuations $Q_\perp$ as a function of wavevector $q$.
    \textbf{(a)} $\zeta=10^{-4}$.
    \textbf{(b)} $\zeta=10^{-1}$.
    In each panel, the curves correspond to a stable state with $|\alpha|=0.08$ (blue), the critical activity $|\alpha|=\alpha_c^I$ (black), and an unstable state with $|\alpha|=0.50$ (red). The corresponding critical activities are $\alpha_c^I=0.124$ in (a) and $\alpha_c^I=0.299$ in (b). The remaining parameters, reported in solver units, are $K=r_0=0.2$, $\eta=\lambda_N=\gamma=1$, $\langle\phi\rangle=1$, and $\phiIN=1.43$, corresponding to $\tilde r=0.06$. Wavevectors are expressed in inverse length units and growth rates in inverse time units. Positive values of $i\omega_\perp$ correspond to unstable modes. Increasing $\zeta$ decreases the screening length $\ell_s=\sqrt{\eta/\zeta}$, which strengthens hydrodynamic screening and therefore lowers the peak and shifts it to higher $q$.
    }
    \label{fig:Linear_stability_trasverse}
\end{figure*}

The instability is controlled entirely by the deviatoric active stress.  The critical value of activity above which the system goes unstable corresponds to  where the maximum of the growth rate intersects the $q$-axis. As a function of $q$, the critical activity takes the form
\begin{equation}
\alpha_c^I(q)
= \,\frac{2}{\lambda_N\,\gamma}\,
\left[
\frac{\zeta\,\tilde r}{q^2}
+ \zeta K
+ \eta\,\tilde r
+ \eta K q^2
\right].
\end{equation}
The maximum of the growth rate is given by 
\begin{align}
&q_0=
 \left[\frac{1}{\eta}\left(
\sqrt{\frac{\lambda_N\,\gamma\,\zeta}{2K}\,|\alpha|}
\;-\; \zeta
\right)\right]^{1/2} ~~~~{\rm for}~~|\alpha|>\frac{2\zeta K}{\lambda_N\gamma}\notag\;,\\
&q_0=0~~~~{\rm for}~~|\alpha|<\frac{2\zeta K}{\lambda_N\gamma}
\end{align}
The threshold for instability is determined by the condition $i\omega_\perp(q_0)=0$. The solution to these equations gives the magnitude of the critical activity above which the uniform quiescent state is unstable as
\begin{equation}
\alpha_c^I(q_c)
= \,\frac{2 \zeta K}{\lambda_N\,\gamma}\;
\left(1 + \sqrt{\frac{\eta\,\tilde r}{\zeta K}}\right)^{2}.
\end{equation}
The condition for the maximum to be on the $q$-axis gives
\begin{equation}
q_c^{\,2} \;=\; \sqrt{\frac{\zeta\,\tilde r}{\eta \, K}}.
\end{equation}

We can write this equation in terms of the relevant time and length-scales of the system: 
\begin{align}
\xi &= \sqrt{\frac{K}{\tilde r}},
& \quad &\text{nematic correlation length}, \notag\\
\ell_s &= \sqrt{\frac{\eta}{\zeta}},
& &\text{flow screening length}, \notag\\
\ell_\alpha &= \sqrt{\frac{\eta K}{\gamma|\alpha|}},
& &\text{active length}, \notag\\
\tau_r &= \frac{\gamma}{\tilde r},
& &\text{nematic relaxation time}.
\end{align}
We then obtain 
\begin{equation}
\boxed{
\alpha_{c,\perp}^I
= \frac{2\zeta K}{\gamma\lambda_N}\;
\left(1 + \frac{\ell_s}{\xi}\right)^{2}.}
\label{eq:alphaI}
\end{equation}
Specializing to extensile systems, $\alpha<0$, the dispersion relation can be written as
\begin{equation}
i \omega_{\perp}(q)\tau_{r}
= - \left[
1 + \xi^2q^2\left(
1
-
\frac{\lambda_N\gamma|\alpha|/(\zeta K)}
{2(1+\ell_s^2 q^2)}
\right)
\right]\;.
\label{eq:stability_perp_dimensionless_SI}
\end{equation}
In the friction-dominated regime, $q\ll\ell_s^{-1}$, this becomes
\begin{equation}
i \omega_{\perp}(q)\tau_{r}
\simeq
-\left[
1+\xi^2q^2
\left(
1-\frac{\lambda_N}{2}\frac{\ell_s^2}{\ell_\alpha^2}
\right)
\right]\;,
\label{eq:fr-inst}
\end{equation}
showing that activity softens the effective elastic constant~\cite{srivastava2016negative}. In the viscosity-dominated regime, $q\gg\ell_s^{-1}$,
\begin{equation}
i \omega_{\perp}(q)\tau_{r}
\simeq
-\left[
\left(
1-\frac{\lambda_N}{2}\frac{\xi^2}{\ell_\alpha^2}
\right)
+\xi^2q^2
\right]\;.
\label{eq:v-inst}
\end{equation}
The homogeneous isotropic state is unstable for  $|\alpha|>\alpha_{c,\perp}^I$ (Eq.~\eqref{eq:alphaI}).

\subsection{Longitudinal component}
\label{LS_longitudinal}

The longitudinal component of the velocity couples density and $Q_\parallel$ fluctuations as
\begin{align}
\partial_t
\begin{pmatrix}
\delta\phi \\[0.5ex]
Q_{\parallel}
\end{pmatrix}
=\mathbf{M}^I
\begin{pmatrix}
\delta\phi \\[0.5ex]
Q_{\parallel}
\end{pmatrix}\;,
\label{eq:matrix}
\end{align}
where
\begin{equation}
\begin{aligned}
\mathbf{M}^I(q)
&=
\begin{pmatrix}
M_{11}^I(q) & M_{12}^I(q)\\
M_{21}^I(q) & M_{22}^I(q)
\end{pmatrix},
\\[1ex]
M_{11}^I(q)
&=
- D q^{2}
-
\frac{q^{2}}
{\kappa_\alpha\zeta\left(1+\ell_s^2q^2\right)},
\\
M_{12}^I(q)
&=
-\frac{|\alpha|}{\zeta}
\frac{q^{2}}{1+\ell_s^2q^2},
\\
M_{21}^I(q)
&=
\frac{\lambda_N q^{2}}
{2\kappa_\alpha\zeta\left(1+\ell_s^2q^2\right)},
\\
M_{22}^I(q)
&=
-\tau_r^{-1}\left(1+\xi^2q^2\right)
+
\frac{\lambda_N|\alpha|}{2\zeta}
\frac{q^{2}}{1+\ell_s^2q^2}.
\end{aligned}
\label{eq:M}
\end{equation}
where we have defined an effective compressibility as
\begin{equation}
\frac{1}{\kappa_\alpha}
=
|\alpha_B|
+
\frac{1}{\kappa}\;.
\end{equation}
The two eigenvalues of the stability matrix are
\begin{equation}
\Lambda_\pm(q)
=
\frac{1}{2}
\left[
\operatorname{Tr}M^I(q)
\pm
\sqrt{\bigl(\operatorname{Tr}M^I(q)\bigr)^2-4\det M^I(q)}
\right]\;,
\label{eq:long-modes}
\end{equation}
with
\begin{equation}
\operatorname{Tr}M^I(q)
=
-\tau_r^{-1}
\left[
1+\left(D\tau_r+\xi^2\right)q^2
\right]
+
\left(
\frac{\lambda_N|\alpha|}{2\zeta}
-
\frac{1}{\zeta\kappa_\alpha}
\right)
\frac{q^2}{1+\ell_s^2q^2}\;,
\end{equation}
and 
\begin{equation}
\det M^I(q)
=
\tau_r^{-1}
\left\{
Dq^2\left(1+\xi^2q^2\right)
+
\frac{q^2}{1+\ell_s^2q^2}
\left[
\frac{1+\xi^2q^2}{\zeta\kappa_\alpha}
-
\frac{D\lambda_N|\alpha|\tau_r}{2\zeta}q^2
\right]
\right\}\;.
\end{equation}

When the discriminant of Eq.~\eqref{eq:long-modes} is negative, the modes form a complex-conjugate pair,
\begin{equation}
\Lambda_\pm(q)
=
\frac{\operatorname{Tr}M^I(q)}{2}
\pm
\frac{i}{2}
\sqrt{4\det M^I(q)-\bigl(\operatorname{Tr}M^I(q)\bigr)^2}\;.
\end{equation}
These modes describe propagating density--nematic waves. A density fluctuation generates a longitudinal flow through the passive and active pressures. Flow alignment then generates a longitudinal nematic fluctuation $Q_\parallel$, whose deviatoric active stress feeds back on the density. Since $M_{12}^I<0$ and $M_{21}^I>0$, this feedback can produce an oscillatory response rather than monotonic relaxation. These are overdamped waves generated by the coupling of density and nematic order, rather than inertial sound waves.

\begin{figure*}[htpb!]
    \centering
    \includegraphics[width=0.98\textwidth]{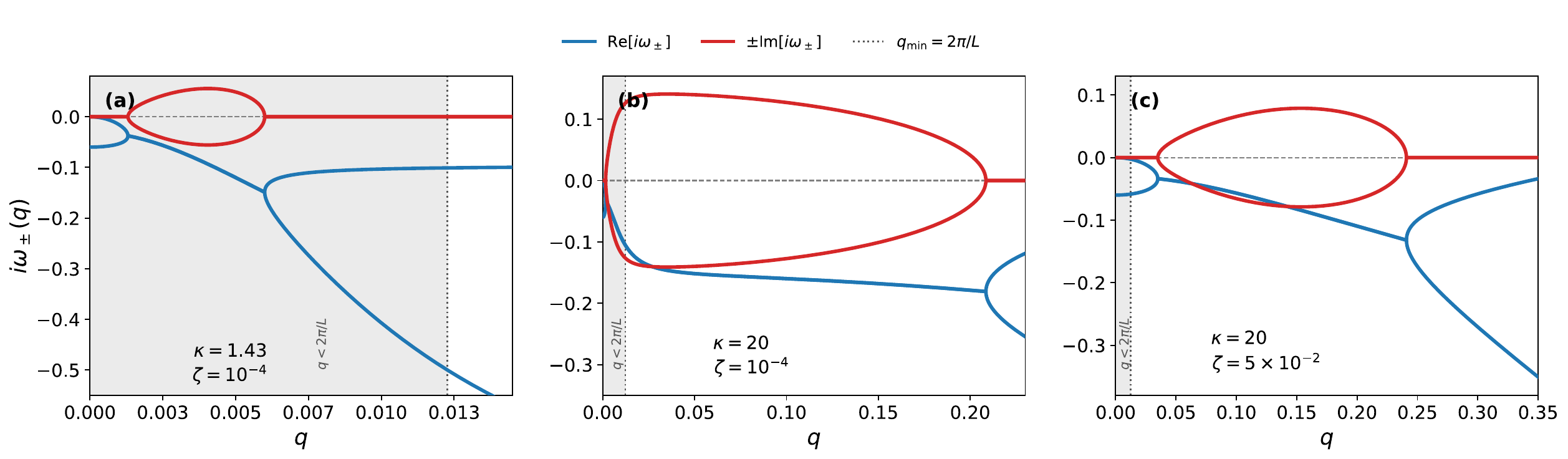}
    \caption{
    \textbf{Dispersion relation for the longitudinal modes.}
    The blue curves show the real parts of the two longitudinal eigenvalues $i\omega_\pm(q)$, while the red curves show their imaginary parts. The shaded region marks wavevectors below the smallest nonzero wavevector $q_{\min}=2\pi/L$ accessible in the finite simulation box.
    \textbf{(a)} For the simulation parameters $\kappa=1.43$ and $\zeta=10^{-4}$, the eigenvalues form a stable complex-conjugate pair over $0.0013\lesssim q\lesssim0.0060$, entirely below $q_{\min}\simeq0.0123$; propagating longitudinal modes are therefore inaccessible in the simulations.
    \textbf{(b)} At the same friction but larger compressibility, $\kappa=20$, the complex-conjugate interval broadens to $0.0016\lesssim q\lesssim0.209$ and includes simulation-accessible wavevectors.
    \textbf{(c)} Same as (b), but with $\zeta=5\times10^{-2}$, illustrating that stronger friction shifts the propagating interval to higher $q$ and reduces the imaginary part of the eigenvalues.
    The remaining parameters, reported in solver units, are $D=1$, $K=r_0=0.2$, $\tilde r=0.06$, $\eta=\gamma=\lambda_N=1$, $\alpha=-1$, $\alpha_B=-0.7$, $\langle\phi\rangle=1$, $\phiIN=1.43$, and $L=512$. Wavevectors are expressed in inverse length units and eigenvalues in inverse time units.
    }
    \label{fig:longitudinal_instability}
\end{figure*}

At small $q$, the density--nematic coupling vanishes as $q^2$, while $Q_\parallel$ retains the finite relaxation rate $\tau_r^{-1}$. At large $q$, density diffusion and nematic elasticity dominate. The propagating waves therefore occur over an intermediate range of wavevectors. Whether this range is accessible also depends on compressibility: at low $\kappa$ it lies below the smallest wavevector of the finite simulation box, whereas at high $\kappa$ it extends to accessible $q$ (Fig.~\ref{fig:longitudinal_instability}). The waves decay when $\operatorname{Tr}M^I(q)<0$ and grow when $\operatorname{Tr}M^I(q)>0$.

When the discriminant is non-negative, the eigenvalues are real. Both modes are stable when
\begin{equation}
\operatorname{Tr}M^I(q)<0,
\qquad
\det M^I(q)>0.
\end{equation}
The real modes are unstable when $\det M^I(q)<0$, in which case one eigenvalue is positive and the other negative, or when $\operatorname{Tr}M^I(q)>0$, in which case at least one eigenvalue is positive.

The longitudinal sector can lose stability in two ways. The maximum of $\operatorname{Tr}M^I(q)$ becomes positive for
\begin{equation}
|\alpha|
>
\frac{2}{\lambda_N\kappa_\alpha}
+
\frac{2\zeta}{\lambda_N\tau_r}
\left(
\sqrt{D\tau_r+\xi^2}+\ell_s
\right)^2.
\end{equation}
If the eigenvalues are complex at this threshold, the propagating waves become unstable. Alternatively, a real eigenvalue becomes unstable when $\det M^I(q)$ crosses zero. Minimizing the corresponding activity over $q$ gives
\begin{equation}
|\alpha|
>
\frac{2\eta}{\lambda_N\tau_r}
\left[
1
+
\sqrt{\frac{\xi^2}{\ell_s^2}
+\frac{\xi^2}{D\eta\kappa_\alpha}}
\right]^2.
\end{equation}
The longitudinal instability threshold is therefore
\begin{equation}
\alpha_{c,\parallel}^I
=
\min
\left\{
\frac{2}{\lambda_N\kappa_\alpha}
+
\frac{2\zeta}{\lambda_N\tau_r}
\left(
\sqrt{D\tau_r+\xi^2}+\ell_s
\right)^2,
\;
\frac{2}{\lambda_N\tau_r}
\left[
\sqrt{\eta}
+
\xi
\sqrt{
\zeta+\frac{1}{D\kappa_\alpha}
}
\right]^2
\right\}.
\label{eq:alpha_c_parallel_I}
\end{equation}

Both possible longitudinal thresholds exceed the transverse threshold $\alpha_{c,\perp}^I$ given in Eq.~\eqref{eq:alphaI}. The first is larger because $\kappa_\alpha^{-1}>0$ and $\sqrt{D\tau_r+\xi^2}>\xi$, while the second is larger because
\begin{equation}
\sqrt{
\zeta+\frac{1}{D\kappa_\alpha}
}
>
\sqrt{\zeta}.
\end{equation}
Thus, as $|\alpha|$ is increased, the transverse instability always occurs first. By the time either longitudinal instability could occur, the homogeneous isotropic state about which it was calculated has already become unstable.

\section{Linear Stability of the Homogeneous Nematic state}

We now examine the linear stability of Eqs.~(1)--(3) in the main text about a homogeneous, ordered nematic state with $\phi=\langle\phi\rangle$, $\mathbf{v}=0$, and director aligned along $\hat{\mathbf{x}}$. We define concentration and orientational fluctuations as $\delta\phi=\phi-\langle\phi\rangle$ and $\delta\theta=\theta-\langle\theta\rangle$, respectively. Since the reference director is aligned along $\hat{\mathbf{x}}$, one has $\langle\theta\rangle=0$. We therefore parametrize the director as $\hat{\mathbf{n}}=(\cos\delta\theta,\sin\delta\theta)$. Deep in the ordered state ($S=1$), fluctuations of the scalar order parameter decay rapidly and are neglected.

To linear order in $\delta\theta\ll1$, the nematic tensor components are $Q_{xx}=\frac{1}{2}$, $Q_{xy}=\delta\theta$, and $Q_{yy}=-\frac{1}{2}$. The angular dynamics then follow from Eq.~(3) in the main text as
\begin{equation}
\partial_t \delta\theta
=
\frac{\lambda_N+1}{2}\,\partial_x v_y
+
\frac{\lambda_N-1}{2}\,\partial_y v_x
+
\frac{K}{\gamma}\nabla^2 \delta\theta.
\label{eq:nematic_angle_linear}
\end{equation}

We analyze fluctuations in Fourier space and introduce the angle $\varphi$ via $\cos\varphi=\hat{\mathbf{x}}\cdot\hat{\mathbf{q}}$. Using the active stress and pressure already defined above, the linearized flow equation can be written directly as
\begin{equation}
v_i(\mathbf q)
=
\frac{i}{\zeta+\eta q^2}
\left[
\alpha q_j\delta Q_{ij}
+
\left(
\alpha Q^0_{ij}
-
\frac{\delta_{ij}}{\kappa_\alpha}
\right)
q_j\delta\phi
\right],
\label{eq:nematic_velocity_linear}
\end{equation}
where we used $|\alpha_B|+1/\kappa=1/\kappa_\alpha$ for extensile isotropic activity.
Substituting Eq.~\eqref{eq:nematic_velocity_linear} into the concentration and angle equations gives
\begin{equation}
\partial_t
\begin{pmatrix}
\delta\phi\\
\delta\theta
\end{pmatrix}
=
\mathbf M^N(q,\varphi)
\begin{pmatrix}
\delta\phi\\
\delta\theta
\end{pmatrix},
\label{eq:nematic_matrix_dynamics}
\end{equation}
with
\begin{equation}
\mathbf M^N(q,\varphi)
=
\begin{pmatrix}
M_{11}^N(q,\varphi) & M_{12}^N(q,\varphi)\\
M_{21}^N(q,\varphi) & M_{22}^N(q,\varphi)
\end{pmatrix},
\end{equation}
where
\begin{align}
M_{11}^N(q,\varphi)
&=
-Dq^2
-
\frac{q^2}{\zeta(1+\ell_s^2q^2)}
\left(
\frac{1}{\kappa_\alpha}
-
\frac{\alpha}{2}\cos2\varphi
\right)\;,
\nonumber\\
M_{12}^N(q,\varphi)
&=
\frac{\alpha q^2}{\zeta(1+\ell_s^2q^2)}
\sin2\varphi\;,
\nonumber\\
M_{21}^N(q,\varphi)
&=
\frac{q^2}{2\zeta(1+\ell_s^2q^2)}
\left(
\frac{\lambda_N}{\kappa_\alpha}
+
\frac{\alpha}{2}
\right)
\sin2\varphi\;,
\nonumber\\
M_{22}^N(q,\varphi)
&=
-\frac{K}{\gamma}q^2
-
\frac{\alpha q^2}{2\zeta(1+\ell_s^2q^2)}
\left(
\lambda_N+\cos2\varphi
\right)\;.
\label{eq:nematic_matrix_elements}
\end{align}

For the special directions $\varphi=0$ and $\varphi=\pi/2$, bend and splay decouple from density fluctuations because $\sin2\varphi=0$. In these directions, the orientational instability is therefore identical to that of an incompressible active nematic, with the familiar extensile bend and contractile splay instabilities~\cite{doostmohammadi2018active,shankar2019hydrodynamics,giomi2015geometry}. The corresponding director growth rates are
\begin{align}
i\omega_{\rm bend}(q)
&=
-\alpha
\frac{\lambda_N+1}{2}
\frac{q^2}{\eta q^2+\zeta}
-
\frac{K}{\gamma}q^2\;,
\nonumber\\
i\omega_{\rm splay}(q)
&=
-\alpha
\frac{\lambda_N-1}{2}
\frac{q^2}{\eta q^2+\zeta}
-
\frac{K}{\gamma}q^2\;.
\label{eq:nematic_bend_splay_growth}
\end{align}
Among these two decoupled directions, the bend mode has the lower threshold for extensile systems with $\lambda_N>0$. A direct analysis of the full angular-dependent stability matrix shows that, for the parameter regime considered here, no oblique mode becomes unstable below the bend threshold; the first instability of the homogeneous nematic state is therefore the pure bend mode.

Setting $i\omega_{\rm bend}(q)=0$ gives the positive critical activity magnitude
\begin{equation}
\alpha_c^N(q)
=
\frac{2K}{\gamma(\lambda_N+1)}
\left(\eta q^2+\zeta\right).
\end{equation}
The most unstable wavevector is
\begin{equation}
q_*^2(\alpha)
=
\frac{1}{\eta}
\left[
\sqrt{
\frac{|\alpha|\gamma\zeta(\lambda_N+1)}
{2K}
}
-\zeta
\right].
\end{equation}
At the onset of instability, $q_*^2=0$, giving
\begin{equation}
\boxed{
\alpha_c^N
=
\frac{2\zeta K}{\gamma(\lambda_N+1)}\;.
}
\end{equation}

\section{Density fluctuations}

\begin{figure*}[htpb!]
\centering
\includegraphics[width=\columnwidth]{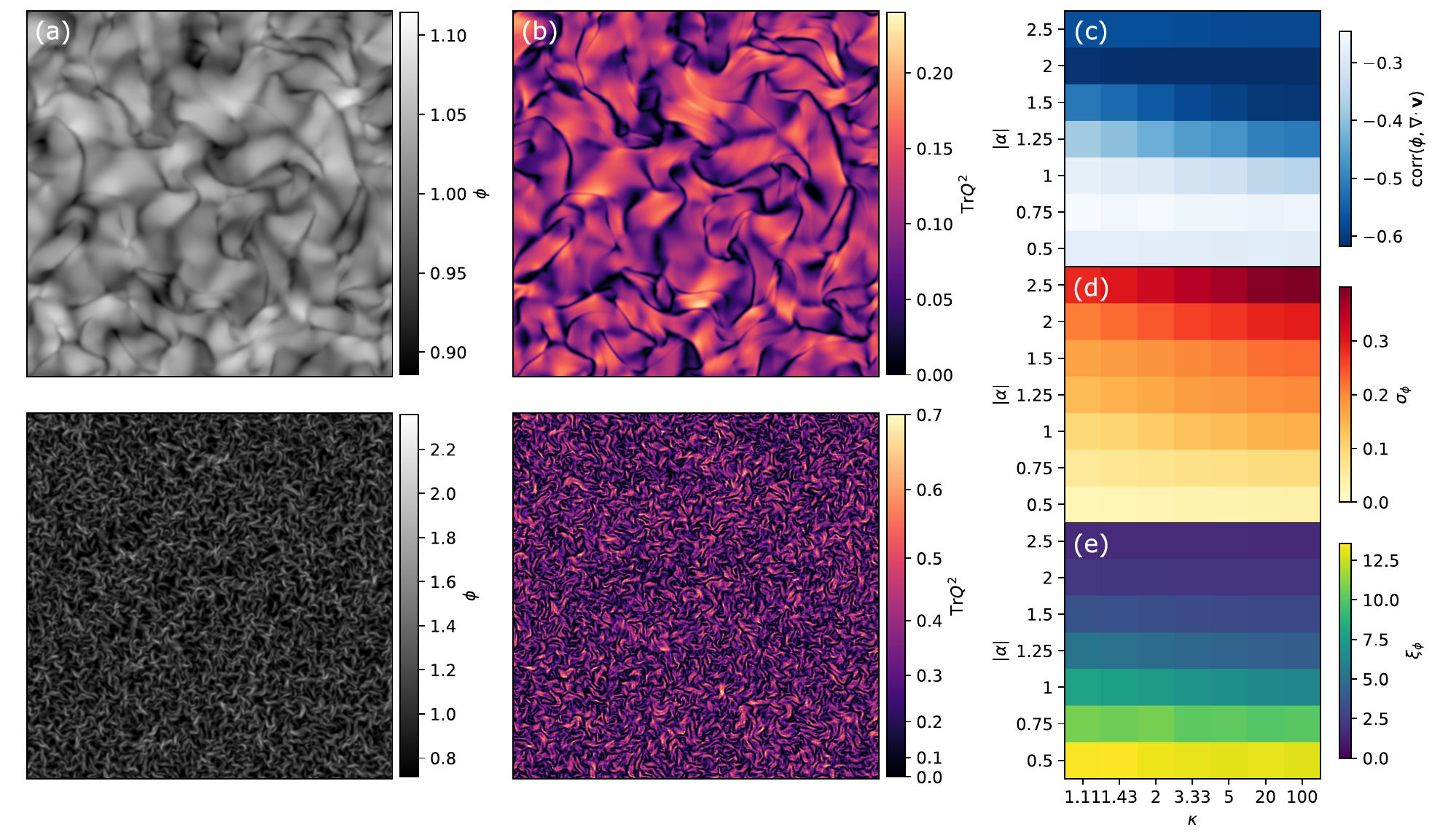}
\caption{
\textbf{Concentration fluctuations and real-space structure in the pure-viscous active turbulent regime.}
Simulations are performed at $\zeta=0$ and fixed $\alpha_B=-1.5$.
Panels (a,b) show representative real-space snapshots of the normalized nematogen concentration $\phi=\rho/\langle\rho\rangle$ and nematic order $\mathrm{Tr}Q^2$ at $\kappa=1.43$ for $|\alpha|=0.5$ (top row) and $|\alpha|=1.5$ (bottom row).
At lower activity, the system forms broad, long-correlated concentration modulations with relatively small amplitude. At higher activity, the concentration field breaks into sharper dense and dilute regions, with enhanced nematic order inside the dense clumps. Separate color ranges are used for the two rows to resolve the spatial structure in both activity regimes.
Panels (c--e) show parameter-space maps over the $7\times7$ sweep in compressibility $\kappa$ and extensile-activity magnitude $|\alpha|$.
Panel (c) shows the pointwise Pearson correlation $\mathrm{corr}(\phi,\nabla\!\cdot\mathbf{v})$. Its negative value associates concentration-rich regions with convergent flow and demonstrates that the clumps are generated by the compressive contribution $-\phi\,\nabla\!\cdot\mathbf{v}$ to the concentration dynamics.
Panel (d) shows the rms concentration fluctuation $\sigma_\phi=[\langle(\phi-\langle\phi\rangle)^2\rangle]^{1/2}$.
Panel (e) shows the concentration correlation length $\xi_\phi$, defined by the first $1/e$ crossing of the radial autocorrelation of $\phi$.
Increasing activity therefore produces stronger but more spatially localized concentration inhomogeneities, while increasing $\kappa$ generally increases their amplitude and decreases their correlation length.
}
\label{fig:zeta0_density_summary}
\end{figure*}

We first characterize the density fluctuations observed numerically under uniform activity in the pure-viscous regime, $\zeta=0$ (Fig.~\ref{fig:zeta0_density_summary}). At fixed $\kappa=1.43$, lower activity produces broad, long-correlated concentration modulations of relatively small amplitude [Fig.~\ref{fig:zeta0_density_summary}(a,b), top]. As activity increases, these structures are replaced by sharper, high-concentration clumps separated by dilute regions [Fig.~\ref{fig:zeta0_density_summary}(a), bottom]. Because the liquid crystal is lyotropic, the dense clumps develop enhanced local nematic order [Fig.~\ref{fig:zeta0_density_summary}(b), bottom].

The parameter-space maps in Fig.~\ref{fig:zeta0_density_summary}(c--e) show how this crossover depends on both activity and $\kappa$. The pointwise correlation $\mathrm{corr}(\phi,\nabla\!\cdot\mathbf{v})$ is negative throughout the parameter range and becomes increasingly negative as $|\alpha|$ is increased [Fig.~\ref{fig:zeta0_density_summary}(c)]. This identifies compressive flow as the mechanism responsible for the concentration inhomogeneities. Indeed, expanding the continuity equation gives
\begin{equation}
    \partial_t\phi+\mathbf{v}\cdot\nabla\phi
    =
    -\phi\,\nabla\!\cdot\mathbf{v}
    +D\nabla^2\phi\;,
\end{equation}
so convergent regions, where $\nabla\!\cdot\mathbf{v}<0$, locally increase $\phi$, while divergent regions deplete it. Increasing activity therefore strengthens the coupling between concentration and compressive flow, producing larger-amplitude and more localized concentration inhomogeneities. Correspondingly, the rms concentration fluctuation $\sigma_\phi$ increases strongly with $|\alpha|$ [Fig.~\ref{fig:zeta0_density_summary}(d)], while the concentration correlation length $\xi_\phi$ decreases [Fig.~\ref{fig:zeta0_density_summary}(e)]. At fixed activity, increasing $\kappa$ generally increases $\sigma_\phi$ and decreases $\xi_\phi$, although this dependence is weaker than the dependence on activity.

We now ask whether these nonlinear concentration structures also produce anomalous long-wavelength number fluctuations. To address this question, we calculate the density structure factor and number fluctuations about homogeneous isotropic and nematic states~\cite{simha2003active,mishra2006active,marchetti2013hydrodynamics,baskaran2008hydrodynamics}.

\subsection{Density fluctuations about the isotropic state}
\label{sec:density_fluctuations_isotropic_state}

We now calculate density correlations about the homogeneous isotropic state. The linearized deterministic dynamics of the coupled density and longitudinal tensor fluctuations, together with the definitions of all relevant quantities, were derived in Sec.~\ref{LS_longitudinal}. In particular, the dynamics are governed by the stability matrix $\mathbf M^I(q)$ defined in Eq.~\eqref{eq:M}. To calculate density correlation functions, we now add conserved density noise $\xi_\phi$ and nematic noise $\xi_L$ to the same coupled longitudinal block:
\begin{equation}
    \partial_t
    \begin{pmatrix}
        \delta\phi\\
        Q_{\parallel}
    \end{pmatrix}
    =
    \mathbf{M}^I(q)
    \begin{pmatrix}
        \delta\phi\\
        Q_{\parallel}
    \end{pmatrix}
    +
    \begin{pmatrix}
        \xi_\phi\\
        \xi_L
    \end{pmatrix}.
    \label{eq:iso_matrix}
\end{equation}
Thus, the deterministic matrix elements $M_{ij}^I(q)$ and eigenvalues $\Lambda_\pm(q)$ are exactly those obtained in the linear-stability analysis.

In frequency space,
\begin{equation}
    \mathcal L_I(q,\omega)
    \begin{pmatrix}
        \delta\phi\\
        Q_{\parallel}
    \end{pmatrix}
    =
    \begin{pmatrix}
        \xi_\phi\\
        \xi_L
    \end{pmatrix},
    \qquad
    \mathcal L_I(q,\omega)
    =
    -i\omega\mathbf{1}-\mathbf{M}^I(q)
    =
    \begin{pmatrix}
        -i\omega-M_{11}^I & -M_{12}^I\\
        -M_{21}^I & -i\omega-M_{22}^I
    \end{pmatrix}.
\end{equation}
Using the eigenvalues $\Lambda_\pm(q)$ of $\mathbf M^I(q)$, the determinant can be written as
\begin{equation}
\mathcal D_I(q,\omega)
=
\det\mathcal L_I(q,\omega)
=
\left(i\omega+\Lambda_+\right)
\left(i\omega+\Lambda_-\right).
\end{equation}
We take
\begin{equation}
    \left\langle
        \xi_\phi(\mathbf q,\omega)
        \xi_\phi(-\mathbf q,-\omega)
    \right\rangle
    =
    2\Delta_\phi q^2,
    \qquad
    \left\langle
        \xi_L(\mathbf q,\omega)
        \xi_L(-\mathbf q,-\omega)
    \right\rangle
    =
    2\Delta_Q\;,
\end{equation}
with no cross-correlations. The dynamic density structure factor is then
\begin{equation}
    S_{\phi\phi}^{I}(q,\omega)
    =
    \frac{
        2\Delta_\phi q^2\left(\omega^2+\left(M_{22}^I\right)^2\right)
        +
        2\Delta_Q \left(M_{12}^I\right)^2
    }{
        \left|
\left(i\omega+\Lambda_+\right)
\left(i\omega+\Lambda_-\right)
\right|^2
    }\;.
\label{eq:iso_Sphiphi_omega}
\end{equation}

The equal-time density structure factor is
\begin{equation}
    S_\phi^I(q)
    =
    \int_{-\infty}^{\infty}
    \frac{d\omega}{2\pi}
    S_{\phi\phi}^{I}(q,\omega).
\end{equation}
The tensor-noise contribution is
\begin{equation}
    S_\phi^{I,(Q)}(q)
    =
    -\frac{
        \Delta_Q \left(M_{12}^I(q)\right)^2
    }{
        \left[\Lambda_+(q)+\Lambda_-(q)\right]
        \Lambda_+(q)\Lambda_-(q)
    }\;.
    \label{eq:iso_equal_time_Q}
\end{equation}
Since $Q_{\parallel}$ is massive in the isotropic state,
\begin{equation}
    M_{22}^I(0)=-\tau_r^{-1}<0\;,
\end{equation}
and the tensor-noise contribution is therefore analytic at small $q$. The conserved density noise also gives a regular small-$q$ contribution. Hence, for a stable isotropic state,
\begin{equation}
    S_\phi^I(q\to0)\sim q^0\;.
    \label{eq:iso_q0_result}
\end{equation}
This conclusion is unchanged by the Brinkman crossover. In the friction-dominated regime, $\zeta+\eta q^2\simeq\zeta$, while in the pure-viscous regime, $\zeta+\eta q^2\simeq\eta q^2$. In both cases, as long as the isotropic state remains linearly stable, the equal-time density structure factor remains finite as $q\to0$:
\begin{equation}
    S_\phi^I(q\to0)\sim q^0\;,
    \qquad
    S_\rho^I(q\to0)=\rho_0^2S_\phi^I(q\to0)\sim q^0\;.
\end{equation}
Thus the isotropic state has normal number fluctuations,
\begin{equation}
    \Delta N\sim N^{1/2}\;.
\end{equation}

The stable isotropic state therefore does not exhibit giant number fluctuations. The reason is that the tensor field $Q_{ij}$ is massive at $q=0$, so tensor fluctuations do not provide a soft Goldstone mode capable of generating a $q^{-2}$ density structure factor. The isotropic calculation is useful for understanding the onset of the active instability, but it does not produce the Simha--Ramaswamy giant-number-fluctuation mechanism~\cite{simha2002hydrodynamic}, which requires a locally ordered nematic state with a soft director mode.

\textit{Caveat:} There is one useful caveat in the pure-viscous limit. For $\zeta\to0$ at fixed $\eta$,
\begin{equation}
    \zeta+\eta q^2\simeq\eta q^2\;.
\end{equation}
The small-$q$ matrix elements then approach
\begin{equation}
    M_{11}^I(q)
    =
    -\frac{1}{\eta\kappa_\alpha}
    -
    Dq^2\;,
    \qquad
    M_{12}^I(q)
    =
    \frac{\alpha}{\eta}\;,
\end{equation}
\begin{equation}
    M_{21}^I(q)
    =
    \frac{\lambda_N}{2\eta\kappa_\alpha}\;,
    \qquad
    M_{22}^I(q)
    =
    -\tau_r^{-1}
    -
    \frac{\lambda_N\alpha}{2\eta}
    -
    \frac{K}{\gamma}q^2\;.
\end{equation}
Thus the tensor-noise contribution to the equal-time density structure factor remains
\begin{equation}
    S_\phi^{I,(Q)}(q)
    =
    -\frac{
        \Delta_Q \left(M_{12}^I(q)\right)^2
    }{
        \left[\Lambda_+(q)+\Lambda_-(q)\right]
        \Lambda_+(q)\Lambda_-(q)
    }\;.
\end{equation}
For generic values of $|\alpha|$, this is finite as $q\to0$. Indeed,
\begin{equation}
    M_{11}^I(q)M_{22}^I(q)-M_{12}^I(q)M_{21}^I(q)
    =
    \frac{1}{\eta\kappa_\alpha\tau_r}
    +
    O(q^2)\;,
\end{equation}
so the determinant of the coupled density--tensor block remains nonzero.

The only singular case occurs when the trace of the small-$q$ longitudinal matrix vanishes,
\begin{equation}
    M_{11}^I(0)+M_{22}^I(0)=0.
\end{equation}
Using the signed convention $\alpha<0$, this occurs at
\begin{equation}
    |\alpha|
    =
    \frac{2}{\lambda_N}
    \left(
        \frac{1}{\kappa_\alpha}
        +
        \frac{\eta}{\tau_r}
    \right)\;,
\end{equation}
for $\lambda_N>0$. Equivalently,
\begin{equation}
    M_{11}^I(q)+M_{22}^I(q)
    =
    -\left(D+\frac{K}{\gamma}\right)q^2\;,
\end{equation}
at this critical activity. Since the determinant remains finite,
\begin{equation}
    M_{11}^I(q)M_{22}^I(q)-M_{12}^I(q)M_{21}^I(q)
    =
    \frac{1}{\eta\kappa_\alpha\tau_r}
    +
    O(q^2)\;,
\end{equation}
the equal-time density structure factor scales as
\begin{equation}
    S_\phi^{I,(Q)}(q)
    \sim
    \frac{1}{q^2}\;.
\end{equation}
This is precisely the pure-viscous limit of the longitudinal critical softening already found in the linear-stability analysis of Sec.~\ref{LS_longitudinal}; it is not a distinct instability. It is also not a Goldstone mode, but rather a critical soft mode of the isotropic state.

In the same pure-viscous limit, the transverse mode softens already at
\begin{equation}
    |\alpha|
    =
    \frac{2\eta}{\lambda_N\tau_r}\;.
\end{equation}
For the extensile instability considered here, with $\lambda_N>0$ and a stable compressive response $\kappa_\alpha^{-1}>0$,
\begin{equation}
    \frac{2}{\lambda_N}
    \left(
        \frac{1}{\kappa_\alpha}
        +
        \frac{\eta}{\tau_r}
    \right)
    >
    \frac{2\eta}{\lambda_N\tau_r}\;.
\end{equation}
Thus the $q^{-2}$ singularity of the isotropic density structure factor is a formal feature of the longitudinal critical point, but it is generically preempted by the transverse instability. In practice, the homogeneous isotropic state first loses stability through the transverse active-nematic mode and enters the active turbulent regime before the longitudinal density--tensor critical point is reached.

\subsection{Density fluctuations about the nematic state}
\label{sec:density_fluctuations_nematic_state}

We now calculate density correlations about the homogeneous ordered nematic state. The deterministic linear dynamics of the coupled density and director fluctuations, together with the definitions of $\delta\phi$, $\delta\theta$, $\varphi$, and the stability matrix $\mathbf M^N(q,\varphi)$, were derived in the linear-stability analysis above. To calculate density correlation functions, we now add conserved density noise $\xi_\phi$ and director noise $\xi_\theta$ to the same coupled density--director block:
\begin{equation}
    \partial_t
    \begin{pmatrix}
        \delta\phi\\
        \delta\theta
    \end{pmatrix}
    =
    \mathbf M^N(q,\varphi)
    \begin{pmatrix}
        \delta\phi\\
        \delta\theta
    \end{pmatrix}
    +
    \begin{pmatrix}
        \xi_\phi\\
        \xi_\theta
    \end{pmatrix}\;.
    \label{eq:gnf_matrix}
\end{equation}
Thus, the deterministic matrix elements $M_{ij}^N(q,\varphi)$ are exactly those defined in Eq.~\eqref{eq:nematic_matrix_elements}. In particular, the off-diagonal element $M_{12}^N$, which couples director fluctuations to density fluctuations, is the active curvature-current coupling. It is proportional to the deviatoric active stress, while the isotropic active pressure enters through $\kappa_\alpha$.

The two eigenvalues of $\mathbf M^N(q,\varphi)$ are
\begin{equation}
    \Lambda_\pm^N(\mathbf q)
    =
    \frac{1}{2}
    \left[
        M_{11}^N+M_{22}^N
        \pm
        \sqrt{
            \left(M_{11}^N-M_{22}^N\right)^2
            +
            4M_{12}^NM_{21}^N
        }
    \right]\;.
\end{equation}
In frequency space,
\begin{equation}
    \mathcal L_N(\mathbf q,\omega)
    \begin{pmatrix}
        \delta\phi\\
        \delta\theta
    \end{pmatrix}
    =
    \begin{pmatrix}
        \xi_\phi\\
        \xi_\theta
    \end{pmatrix},
    \qquad
    \mathcal L_N(\mathbf q,\omega)
    =
    -i\omega\mathbf 1-\mathbf M^N
    =
    \begin{pmatrix}
        -i\omega-M_{11}^N & -M_{12}^N\\
        -M_{21}^N & -i\omega-M_{22}^N
    \end{pmatrix}\;.
\end{equation}
The determinant can then be written as
\begin{equation}
    \mathcal D_N(\mathbf q,\omega)
    =
    \det\mathcal L_N(\mathbf q,\omega)
    =
    \left(i\omega+\Lambda_+^N\right)
    \left(i\omega+\Lambda_-^N\right)\;.
\end{equation}
We take
\begin{equation}
    \left\langle
        \xi_\phi(\mathbf q,\omega)
        \xi_\phi(-\mathbf q,-\omega)
    \right\rangle
    =
    2\Delta_\phi q^2,
    \qquad
    \left\langle
        \xi_\theta(\mathbf q,\omega)
        \xi_\theta(-\mathbf q,-\omega)
    \right\rangle
    =
    2\Delta_\theta,
\end{equation}
with no cross-correlations. The dynamic structure-factor matrix elements are
\begin{equation}
    S_{\phi\phi}^N(\mathbf q,\omega)
    =
    \frac{
        2\Delta_\phi q^2
        \left[
            \omega^2+\left(M_{22}^N\right)^2
        \right]
        +
        2\Delta_\theta\left(M_{12}^N\right)^2
    }{
        \left|
            \left(i\omega+\Lambda_+^N\right)
            \left(i\omega+\Lambda_-^N\right)
        \right|^2
    },
    \label{eq:gnf_Sphiphi_omega}
\end{equation}
\begin{equation}
    S_{\theta\theta}^N(\mathbf q,\omega)
    =
    \frac{
        2\Delta_\phi q^2\left(M_{21}^N\right)^2
        +
        2\Delta_\theta
        \left[
            \omega^2+\left(M_{11}^N\right)^2
        \right]
    }{
        \left|
            \left(i\omega+\Lambda_+^N\right)
            \left(i\omega+\Lambda_-^N\right)
        \right|^2
    },
    \label{eq:gnf_Sthetatheta_omega}
\end{equation}
\begin{equation}
    S_{\phi\theta}^N(\mathbf q,\omega)
    =
    \frac{
        2\Delta_\phi q^2 M_{21}^N
        \left(-i\omega-M_{22}^N\right)
        +
        2\Delta_\theta M_{12}^N
        \left(i\omega-M_{11}^N\right)
    }{
        \left|
            \left(i\omega+\Lambda_+^N\right)
            \left(i\omega+\Lambda_-^N\right)
        \right|^2
    },
    \label{eq:gnf_Sphitheta_omega}
\end{equation}
and
\begin{equation}
    S_{\theta\phi}^N(\mathbf q,\omega)
    =
    \frac{
        2\Delta_\phi q^2 M_{21}^N
        \left(i\omega-M_{22}^N\right)
        +
        2\Delta_\theta M_{12}^N
        \left(-i\omega-M_{11}^N\right)
    }{
        \left|
            \left(i\omega+\Lambda_+^N\right)
            \left(i\omega+\Lambda_-^N\right)
        \right|^2
    }.
    \label{eq:gnf_Sthetaphi_omega}
\end{equation}
The singular contribution to the density structure factor comes from the director noise,
\begin{equation}
    S_{\phi\phi}^{N,(\theta)}(\mathbf q,\omega)
    =
    \frac{
        2\Delta_\theta
        \left(M_{12}^N(\mathbf q)\right)^2
    }{
        \left|
            \left(i\omega+\Lambda_+^N\right)
            \left(i\omega+\Lambda_-^N\right)
        \right|^2
    }.
    \label{eq:gnf_dynamic_theta}
\end{equation}

\begin{figure*}[htpb!]
\centering
\includegraphics[width=\columnwidth]{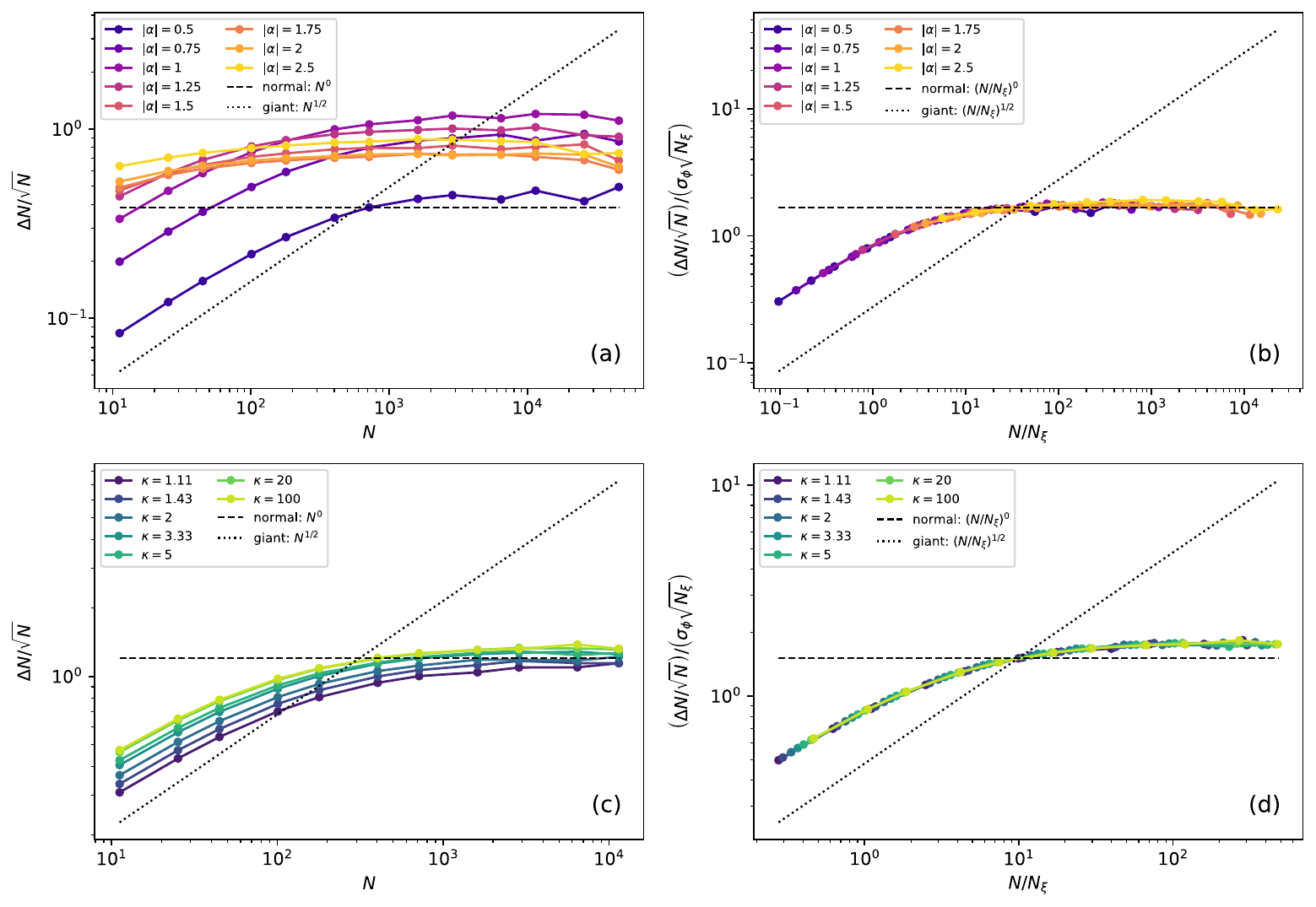}
\caption{
\textbf{Number fluctuations across activity and compressibility in the pure-viscous regime.}
Simulations are performed at $\zeta=0$ and fixed $\alpha_B=-1.5$.
In the top row, $\kappa=1.43$ is fixed while the extensile-activity magnitude $|\alpha|$ is varied.
In the bottom row, $\alpha=-1$ is fixed while $\kappa$ is varied.
Panels (a,c) show $\Delta N/\sqrt{N}$ versus the mean number $N$ in square counting boxes. Normal number fluctuations approach a constant plateau, whereas giant number fluctuations grow as $N^{1/2}$.
Panels (b,d) show the corresponding data rescaled as
$\left(\Delta N/\sqrt{N}\right)/\left(\sigma_\phi\sqrt{N_\xi}\right)$ versus $N/N_\xi$.
Here, $\phi=\rho/\langle\rho\rangle$, $\sigma_\phi$ is the rms fluctuation of $\phi$, $\xi_\phi$ is the correlation length extracted from the radial autocorrelation of $\phi$, and $N_\xi$ is the mean number contained in a correlation area of size $\xi_\phi^2$.
The dashed and dotted lines indicate the normal- and giant-number-fluctuation scalings, respectively.
The collapse of the rescaled curves onto a common large-$N$ plateau shows that the differences between the unscaled curves arise predominantly from changes in the fluctuation amplitude and correlation area, while the asymptotic number fluctuations remain approximately normal.
}
\label{fig:gnf_sqrtN_scaled_zeta0}
\end{figure*}

The equal-time density structure factor is
\begin{equation}
    S_\phi^N(\mathbf q)
    =
    \int_{-\infty}^{\infty}
    \frac{d\omega}{2\pi}
    S_{\phi\phi}^N(\mathbf q,\omega).
\end{equation}
The director-noise contribution is
\begin{equation}
    S_\phi^{N,(\theta)}(\mathbf q)
    =
    -\frac{
        \Delta_\theta
        \left(M_{12}^N(\mathbf q)\right)^2
    }{
        \left[
            \Lambda_+^N(\mathbf q)+\Lambda_-^N(\mathbf q)
        \right]
        \Lambda_+^N(\mathbf q)
        \Lambda_-^N(\mathbf q)
    }.
    \label{eq:gnf_equal_time_theta}
\end{equation}
This is the contribution relevant for giant number fluctuations. The density-noise contribution is regular at small $q$ because the density noise is conserved. The physical density structure factor is related to the normalized-density structure factor by
\begin{equation}
    S_\rho^N(\mathbf q)
    =
    \rho_0^2S_\phi^N(\mathbf q),
\end{equation}
so the scaling with $q$ is unchanged.

\paragraph{Friction-dominated limit.}
For $\eta\to0$ at fixed $\zeta$,
\begin{equation}
    \zeta+\eta q^2\simeq \zeta.
\end{equation}
In this limit, the matrix elements become
\begin{equation}
    M_{11}^N(\mathbf q)
    =
    -q^2
    \left[
        D
        +
        \frac{1}{\zeta}
        \left(
            \frac{1}{\kappa_\alpha}
            -
            \frac{\alpha}{2}\cos2\varphi
        \right)
    \right],
\end{equation}
\begin{equation}
    M_{12}^N(\mathbf q)
    =
    q^2
    \frac{\alpha}{\zeta}
    \sin2\varphi,
\end{equation}
\begin{equation}
    M_{21}^N(\mathbf q)
    =
    q^2
    \frac{1}{2\zeta}
    \left[
        \frac{\lambda_N}{\kappa_\alpha}
        +
        \frac{\alpha}{2}
    \right]
    \sin2\varphi,
\end{equation}
and
\begin{equation}
    M_{22}^N(\mathbf q)
    =
    -q^2
    \left[
        \frac{K}{\gamma}
        +
        \frac{\alpha}{2\zeta}
        \left(
            \lambda_N+\cos2\varphi
        \right)
    \right].
\end{equation}
Substituting these expressions into Eq.~\eqref{eq:gnf_equal_time_theta} gives
\begin{equation}
\begin{aligned}
    S_\phi^{N,(\theta)}(\mathbf q)
    &=
    \frac{1}{q^2}
    \frac{
        \alpha^2\zeta\sin^2 2\varphi\,\Delta_\theta
    }{
        \left[
            \zeta\left(D+\frac{K}{\gamma}\right)
            +
            \frac{1}{\kappa_\alpha}
            +
            \frac{\alpha\lambda_N}{2}
        \right]
    }
    \\
    &\hspace{1.2cm}\times
    \frac{1}{
        \left[
            \zeta D+\frac{1}{\kappa_\alpha}
            -
            \frac{\alpha}{2}\cos2\varphi
        \right]
        \left[
            \frac{\zeta K}{\gamma}
            +
            \frac{\alpha}{2}
            \left(
                \lambda_N+\cos2\varphi
            \right)
        \right]
        -
        \frac{\alpha}{2}
        \left[
            \frac{\lambda_N}{\kappa_\alpha}
            +
            \frac{\alpha}{2}
        \right]
        \sin^2 2\varphi
    }.
\end{aligned}
\label{eq:gnf_friction_result}
\end{equation}
Thus, in the friction-dominated limit,
\begin{equation}
    S_\phi^{N,(\theta)}(q)\sim q^{-2}.
\end{equation}
In two dimensions, this gives the dry active-nematic giant-number-fluctuation result~\cite{simha2002hydrodynamic},
\begin{equation}
    \Delta N\sim N.
\end{equation}

\paragraph{Pure-viscous limit.}
For $\zeta\to0$ at fixed $\eta$,
\begin{equation}
    \zeta+\eta q^2\simeq\eta q^2.
\end{equation}
In this limit,
\begin{equation}
    M_{11}^N(\mathbf q)
    \to
    -\frac{1}{\eta}
    \left[
        \frac{1}{\kappa_\alpha}
        -
        \frac{\alpha}{2}\cos2\varphi
    \right],
\end{equation}
\begin{equation}
    M_{12}^N(\mathbf q)
    \to
    \frac{\alpha}{\eta}
    \sin2\varphi,
\end{equation}
\begin{equation}
    M_{21}^N(\mathbf q)
    \to
    \frac{1}{2\eta}
    \left[
        \frac{\lambda_N}{\kappa_\alpha}
        +
        \frac{\alpha}{2}
    \right]
    \sin2\varphi,
\end{equation}
and
\begin{equation}
    M_{22}^N(\mathbf q)
    \to
    -\frac{\alpha}{2\eta}
    \left[
        \lambda_N+\cos2\varphi
    \right].
\end{equation}
Substituting these expressions into Eq.~\eqref{eq:gnf_equal_time_theta} gives
\begin{equation}
\begin{aligned}
    S_\phi^{N,(\theta)}(\mathbf q)
    \to
    \frac{
        \alpha^2\eta\sin^2 2\varphi\,\Delta_\theta
         }{ \left[  \frac{1}{\kappa_\alpha}
            +
            \frac{\alpha\lambda_N}{2}\right]
        \left[
            \frac{1}{\kappa_\alpha}
            -
            \frac{\alpha}{2}\cos2\varphi
        \right]
        \left[
            \frac{\alpha}{2}
            \left(
                \lambda_N+\cos2\varphi
            \right)
        \right]
        -
        \frac{\alpha}{2}
        \left[
            \frac{\lambda_N}{\kappa_\alpha}
            +
            \frac{\alpha}{2}
        \right]
        \sin^2 2\varphi
    }.
\end{aligned}
\label{eq:gnf_viscous_result}
\end{equation}
Therefore,
\begin{equation}
    S_\phi^{N,(\theta)}(q)\sim q^0.
\end{equation}
Thus the strict pure-viscous linear theory cuts off the $q^{-2}$ density divergence.

For finite $\zeta$ and $\eta$, the crossover is controlled by
\begin{equation}
    q_s
    =
    \left(
        \frac{\zeta}{\eta}
    \right)^{1/2}
    =
    \ell_s^{-1}.
\end{equation}
For $q\ll q_s$, the system is friction dominated and recovers $S_\phi^{N,(\theta)}(q)\sim q^{-2}$. For $q\gg q_s$, the system crosses over toward the viscous regime, where the singular density response is suppressed.

\section{Scalar model}
We begin from the continuum equations in the main text,
\begin{align}
&\partial_t \phi + \nabla \cdot (\phi \mathbf{v})
= D \nabla^2 \phi \;, \\
&\eta \nabla^2 \mathbf{v}
+ \nabla \cdot \left(\alpha\, \phi\, \mathbf{Q}\right)
+ \nabla \left[\alpha_B\,\phi\right]
- \frac{1}{\kappa}\nabla\phi
= \zeta \phi \mathbf{v} \;, \\
&\partial_t \mathbf{Q}
+ \textcolor{red}{(\mathbf{v}\cdot\nabla)\mathbf{Q}}
= \lambda_N \mathbf{A}
- [\boldsymbol{\omega},\mathbf{Q}]
+ \frac{1}{\gamma}\mathbf{H} \;.
\label{eq:1D_starting_eqs}
\end{align}

We now consider a one-dimensional coarse-grained description along the
$x$ direction, with $\phi=\phi(x,t)$, $\mathbf{Q}=0$, and
$\mathbf{v}=v(x,t)\,\hat{\mathbf{x}}$. That is, we focus on density
redistribution driven by a prescribed spatially varying isotropic activity
profile and neglect orientational order at this coarse-grained level. We write
\begin{equation}
\alpha_B(x)=-\alpha_B^0 f(x),
\qquad
\alpha_B^0>0,
\end{equation}
where $0\leq f(x)\leq1$ is periodic with period $L$, or with wavelength
$\lambda$ for the profiles considered below, and define
\begin{equation}
D_p=D+\frac{1}{\kappa\zeta},
\qquad
D_a^0=\frac{\alpha_B^0}{\zeta}.
\end{equation}

Under these assumptions, the continuity and force-balance equations reduce to
\begin{align}
\partial_t \phi + \partial_x(\phi v)
&= D\,\partial_x^2 \phi \;,
\label{eq:1D_mass}\\
\eta\,\partial_x^2 v
&=
-\,\partial_x\!\left[\alpha_B(x)\phi\right]
+\frac{1}{\kappa}\,\partial_x \phi
+\zeta\phi\,v \;.
\label{eq:1D_force}
\end{align}

In the regime where viscous stresses are negligible at this coarse-grained
level, we drop the term $\eta\,\partial_x^2 v$. Solving
Eq.~\eqref{eq:1D_force} for $\phi v$ then gives
\begin{equation}
\phi v
=
\frac{1}{\zeta}\,\partial_x\!\left[\alpha_B(x)\phi\right]
-\frac{1}{\kappa\zeta}\,\partial_x \phi.
\label{eq:rho_v_general}
\end{equation}

Substituting this into Eq.~\eqref{eq:1D_mass} yields the advection--diffusion equation
\begin{equation}
\partial_t\phi
=
\partial_x\!\left[
-v_d(x)\phi
+
D_e(x)\partial_x\phi
\right],
\label{eq:adv-diff}
\end{equation}
where
\begin{equation}
v_d(x)=\frac{\partial_x\alpha_B(x)}{\zeta}
=-D_a^0\partial_x f(x),
\qquad
D_e(x)=D_p-\frac{\alpha_B(x)}{\zeta}
=D_p+D_a^0f(x).
\end{equation}
Since $\partial_xD_e=-v_d$, Eq.~\eqref{eq:adv-diff} can equivalently be written as
\begin{equation}
\partial_t\phi
=
\partial_x^2\!\left[D_e(x)\phi(x,t)\right].
\end{equation}
The effective diffusivity is positive and largest in the high-activity regions, while $v_d$ changes sign across the interfaces and drives material from high- to low-activity regions. This spatially varying drift permits a nonuniform zero-flux steady state.

\subsection{Steady state for a general activity profile}

We now solve Eq.~\eqref{eq:adv-diff} at steady state under periodic
boundary conditions on a domain of length $L$, with
$\phi(x+L)=\phi(x)$ and $\alpha_B(x+L)=\alpha_B(x)$.
Equation~\eqref{eq:adv-diff} can be written as a conservation law,
\begin{equation}
\partial_t\phi+\partial_xj=0,
\qquad
j(x)=v_d(x)\phi(x)-D_e(x)\partial_x\phi(x).
\end{equation}
At steady state,
\begin{equation}
\partial_x j(x)=0,
\qquad
j(x)=j_0,
\label{eq:steady_flux_const}
\end{equation}
where $j_0$ is a constant.

For periodic boundary conditions and zero net transport, we take $j_0=0$.
The steady-state equation then reads
\begin{equation}
v_d(x)\phi(x)-D_e(x)\partial_x\phi(x)=0,
\end{equation}
or equivalently
\begin{equation}
\frac{\partial_x\phi}{\phi}
=
\frac{v_d(x)}{D_e(x)}
=
-\frac{\partial_xD_e(x)}{D_e(x)}.
\end{equation}
So,
\begin{equation}
\phi(x)
=
\frac{C}
{D_p-\dfrac{\alpha_B(x)}{\zeta}},
\label{eq:rho_general_solution}
\end{equation}
where the constant $C$ is fixed by mass conservation,
$\frac{1}{L}\int_0^L\phi(x)\,dx=\langle\phi\rangle$. Namely
\begin{equation}
C
=
\frac{\langle\phi\rangle L}
{\displaystyle
\int_0^L
\frac{dx}
{D_p-\dfrac{\alpha_B(x)}{\zeta}}}.
\label{eq:C_general}
\end{equation}

The corresponding velocity is
$v(x)=D\,\partial_x\phi(x)/\phi(x)$. Using
Eq.~\eqref{eq:rho_general_solution},
\begin{equation}
v(x)
=
-\,D\,
\frac{
\partial_x
\left(
D_p-\dfrac{\alpha_B(x)}{\zeta}
\right)}
{D_p-\dfrac{\alpha_B(x)}{\zeta}}
=
D\,
\frac{\partial_x\alpha_B(x)}
{\zeta D_p-\alpha_B(x)}.
\label{eq:v_general_solution}
\end{equation}

Thus, for a general prescribed activity profile, the steady density and
velocity are completely determined by
Eqs.~\eqref{eq:rho_general_solution} and
\eqref{eq:v_general_solution}, together with the periodic boundary
conditions and mass conservation.

Note that the existence of an inhomogeneous steady state is tied to the spatial
variation of $\alpha_B(x)$. If $\partial_x\alpha_B(x)\equiv0$, then
Eq.~\eqref{eq:rho_general_solution} gives a homogeneous steady state.
By contrast, a spatially varying periodic activity profile generates the
nonuniform  steady density profile given by Eq.~\eqref{eq:rho_general_solution}.
For the prescribed profile $\alpha_B(x)=-\alpha_B^0f(x)$, we introduce
\begin{equation}
R
=
\frac{D_a^0}{D_p}
=
\frac{\alpha_B^0/\zeta}
{D+\dfrac{1}{\kappa\zeta}}
=
\frac{\alpha_B^0\kappa}
{1+\kappa\zeta D},
\label{eq:R_def}
\end{equation}
which is the ratio of activity-induced to passive diffusivity.
Using Eq.~\eqref{eq:R_def} and the normalization
$\langle\phi\rangle=1$, Eqs.~\eqref{eq:rho_general_solution} and
\eqref{eq:C_general} give
\begin{equation}
\phi(x)
=
\frac{\displaystyle
\left[
\frac{1}{L}\int_0^L
\frac{dx}{1+Rf(x)}
\right]^{-1}}
{1+Rf(x)}.
\label{eq:rho_general_R}
\end{equation}
Similarly, Eq.~\eqref{eq:v_general_solution} becomes
\begin{equation}
v(x)
=
-\,D\,
\frac{R\,\partial_x f(x)}
{1+R f(x)}.
\label{eq:v_general_R}
\end{equation}
Equation~\eqref{eq:rho_general_R} is the general steady-state solution
for any imposed profile $f(x)$, subject to periodic boundary conditions
and zero net flux.

\subsection{Sinusoidal activity profile}

\begin{figure}[htpb!]
    \centering
    \includegraphics[width=0.55\textwidth]{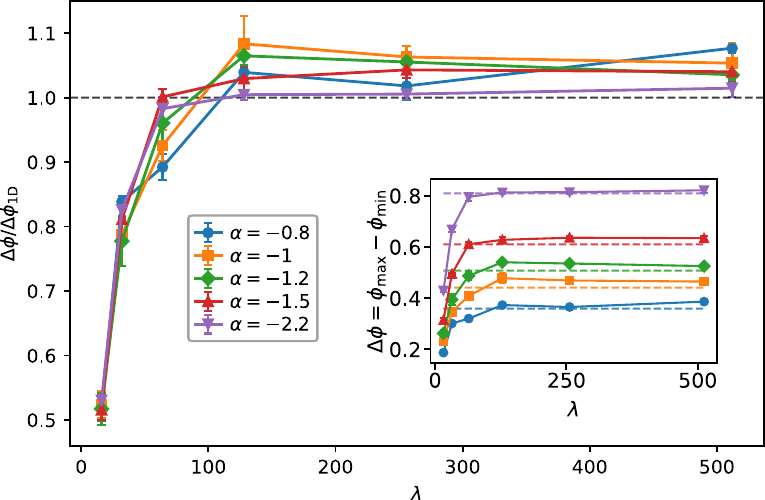}
    \caption{
    Scale-separation test for the one-dimensional isotropic reduction.
    The main panel shows the density contrast measured in the full two-dimensional numerical solutions, $\Delta\phi$, normalized by the one-dimensional prediction $\Delta\phi_{\rm 1D}=R/\sqrt{1+R}$, as a function of the imposed activity wavelength $\lambda$.
    The dashed horizontal line denotes perfect agreement with the one-dimensional theory.
    The inset shows the unscaled contrast $\Delta\phi=\phi_{\max}-\phi_{\min}$; dashed lines indicate the corresponding one-dimensional predictions.
    The parameters are $(\alpha,\alpha_B)=(-0.8,-0.25),(-1.0,-0.32),(-1.2,-0.38),(-1.5,-0.48),(-2.2,-0.70)$.
    The corresponding values are $R=0.429,0.549,0.651,0.823,1.200$ and $\Delta\phi_{\rm 1D}=0.359,0.441,0.507,0.609,0.809$.
    For sufficiently large $\lambda$, the full two-dimensional numerical solutions approach the one-dimensional prediction, while deviations at small $\lambda$ signal the breakdown of the isotropic reduction when the imposed activity varies on scales comparable to the vorticity correlation length $\ell_\omega$.
    }
    \label{fig:si_delta_phi_lambda}
\end{figure}

We first consider
\begin{equation}
f(x)=\frac{1}{2}\left(1+\sin\frac{2\pi x}{\lambda}\right),
\end{equation}
so that $\alpha_B(x)=-\alpha_B^0f(x)$. We take $L/\lambda$ to be an integer, so that the profile is periodic on the domain.
Using Eq.~\eqref{eq:rho_general_R}, the steady-state density profile is
\begin{equation}
\phi(x)
=
\frac{\sqrt{1+R}}
{1+\dfrac{R}{2}\left(1+\sin\dfrac{2\pi x}{\lambda}\right)}.
\label{eq:rho_sine_final}
\end{equation}
From Eq.~\eqref{eq:v_general_R}, the velocity profile is
\begin{equation}
v(x)
=
-\,\frac{D R \pi}{\lambda}\,
\frac{\cos\!\left(\dfrac{2\pi x}{\lambda}\right)}
{1+\dfrac{R}{2}\left(1+\sin\dfrac{2\pi x}{\lambda}\right)}.
\label{eq:v_sine_final}
\end{equation}
The density is maximal where the activity is minimal, namely at $x=\frac{3\lambda}{4}$ (mod $\lambda$), and minimal where the activity is maximal, namely at $x=\frac{\lambda}{4}$ (mod $\lambda$).

Therefore
\begin{equation}
\phi_{\max}=\sqrt{1+R},
\qquad
\phi_{\min}=\frac{1}{\sqrt{1+R}}.
\end{equation}
The density contrast is then
\begin{equation}
\Delta \phi
=
\phi_{\max}-\phi_{\min}
=
\sqrt{1+R}-\frac{1}{\sqrt{1+R}}
=
\boxed{\frac{R}{\sqrt{1+R}}}.
\label{eq:delta_rho_sine}
\end{equation}
To test when this one-dimensional reduction applies to the full two-dimensional dynamics, we compare the measured density contrast in the numerical solutions with the prediction in Eq.~\eqref{eq:delta_rho_sine} while varying the imposed wavelength $\lambda$. As shown in Fig.~\ref{fig:si_delta_phi_lambda}, the contrast approaches the one-dimensional prediction at large $\lambda$, where the activity pattern varies slowly compared to the vorticity correlation length $\ell_\omega$. The numerical solutions used in the main text correspond to $\lambda=L/2$, which falls within this scale-separated regime. The deviations at small $\lambda$ therefore quantify the breakdown of the isotropic reduction when nematic stresses and vortical flow feed back on the density profile.

\REVSTART

\begin{figure}[htpb!]
    \centering
    \includegraphics[
        width=0.7\textwidth
    ]{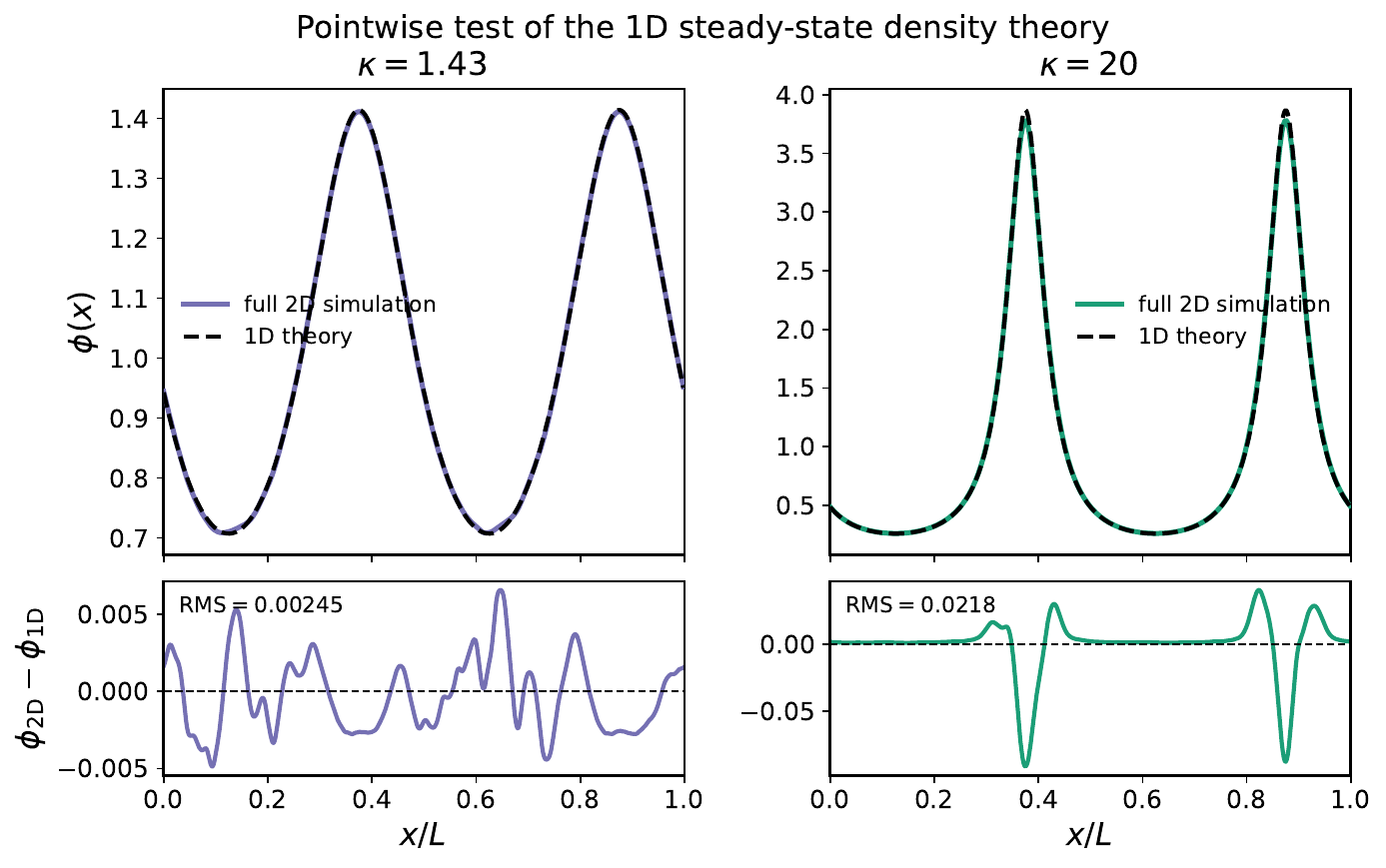}
        \caption{\textcolor{red}{
    \textbf{Pointwise comparison of the steady-state density profiles.}
    Top: concentration profile obtained from the minimal scalar model
    (Eq.~\eqref{eq:rho_sine_final}, dashed line) compared with the profile
    from simulations of the full hydrodynamic model (solid line) for the
    two sinusoidal activity patterns used in Fig.~1 of the main text,
    $\kappa=1.43$ and $\kappa=20$.
    Bottom: pointwise residual
    $\phi_{\rm 2D}-\phi_{\rm 1D}$.
    The numerical profiles are averaged along the direction parallel to the
    activity stripes and over 54 late-time configurations.
    }}
    \label{fig:si_full_phi_profiles}
\end{figure}

We additionally test our prediction for the analytical density profile obtained from the scalar model
[Eq.~\eqref{eq:rho_sine_final}] rather than just the density contrast.
Figure~\ref{fig:si_full_phi_profiles} compares the steady-state
density profiles obtained from simulations of the full two-dimensional hydrodynamic equations with the
one-dimensional analytical result for the two activity patterns shown in
Fig.~1 of the main text, $\kappa=1.43$ and $\kappa=20$. 
In both cases the minimal $1D$ theory
closely reproduces the density profile throughout the domain, showing that
the agreement is not restricted to the extrema used to define
$\Delta\phi$.

\REVEND

\subsection{Square pulse / hyperbolic tangent profile}

We consider the smooth square-pulse profile
\begin{equation}
f(x)
=
\frac{1}{2}\left[
\tanh\!\left(\frac{x-\tfrac{L}{4}}{W}\right)
-
\tanh\!\left(\frac{x-\tfrac{3L}{4}}{W}\right)
\right],
\end{equation}
periodically continued over the domain, so that $\alpha_B(x)=-\alpha_B^0 f(x)$.
Using Eq.~\eqref{eq:rho_general_R}, the steady-state density is
\begin{equation}
\phi(x)
=
\frac{C}{1+Rf(x)},
\end{equation}
where
\begin{equation}
C
=
\left[
\frac{1}{L}
\int_0^L
\frac{dx}{1+Rf(x)}
\right]^{-1}.
\end{equation}
In the limit $W\ll L$, the profile becomes approximately a top-hat:
$f(x)\approx 1$ for $x\in(\tfrac{L}{4},\tfrac{3L}{4})$ and $f(x)\approx 0$ otherwise.
Then
\begin{equation}
\int_0^L \frac{dx}{1+R f(x)}
=
\frac{L}{2}\frac{1}{1+R}+\frac{L}{2},
\end{equation}
so that
\begin{equation}
C
\simeq
\frac{2(1+R)}{2+R}.
\end{equation}
In the sharp-interface limit,
\begin{equation}
\phi(x)
\simeq
\frac{\dfrac{2(1+R)}{2+R}}
{1+\dfrac{R}{2}\left[
\tanh\!\left(\frac{x-\tfrac{L}{4}}{W}\right)
-
\tanh\!\left(\frac{x-\tfrac{3L}{4}}{W}\right)
\right]},
\label{eq:phi-2tanh}
\end{equation}
and
\begin{equation}
v(x)
=
-\,\frac{D R}{2W}
\frac{
\sech^2\!\left(\dfrac{x-\tfrac{L}{4}}{W}\right)
-
\sech^2\!\left(\dfrac{x-\tfrac{3L}{4}}{W}\right)}
{1+\dfrac{R}{2}\left[
\tanh\!\left(\dfrac{x-\tfrac{L}{4}}{W}\right)
-
\tanh\!\left(\dfrac{x-\tfrac{3L}{4}}{W}\right)
\right]}.
\end{equation}
The extrema of the density profile read
\begin{equation}
\phi_{\max}=\frac{2(1+R)}{2+R},
\qquad
\phi_{\min}=\frac{2}{2+R}.
\end{equation}
Thus the density contrast is
\begin{equation}
\Delta\phi
=
\phi_{\max}-\phi_{\min}
=
\frac{2R}{2+R}.
\end{equation}
For $R\ll1$, one has
\begin{equation}
\Delta\phi
=
R+\mathcal{O}(R^2),
\end{equation}
for both the tanh square-pulse and sinusoidal profiles.

\REVSTART

\subsection{Finite-viscosity correction to the scalar steady state}
\label{sec:scalar_finite_viscosity}

The scalar profiles derived above were obtained by neglecting the viscous
term in the one-dimensional force balance. We now retain this term and
quantify its effect specifically for the parameter values of the interfacial vortex state.

At steady state, Eqs.~\eqref{eq:1D_mass} and \eqref{eq:1D_force} give
\begin{align}
\frac{d}{dx}&\left(\phi v-D\phi'\right)=0,
\label{eq:finite_eta_mass}\\
\eta v''
&=
-\frac{d}{dx}\left[\alpha_B(x)\phi\right]
+\frac{1}{\kappa}\phi'
+\zeta\phi v .
\label{eq:finite_eta_force}
\end{align}
For a periodic steady state with no imposed through-flux, the integration
constant in Eq.~\eqref{eq:finite_eta_mass} vanishes, and therefore
\begin{equation}
v
=
D\,\frac{\phi'}{\phi}
=
D\,\partial_x\ln\phi .
\label{eq:finite_eta_velocity}
\end{equation}

Defining
\begin{equation}
y(x)=\ln\phi(x),
\end{equation}
we can reduce 
Eq.~\eqref{eq:finite_eta_force} to 
\begin{equation}
\eta D y'''
=
\frac{d}{dx}
\left\{
\left[
\frac{1}{\kappa}
+\zeta D
-\alpha_B(x)
\right]
e^y
\right\}.
\end{equation}
Integrating once, the finite-viscosity steady state is  determined
by
\begin{equation}
\boxed{
\eta D y''
=
\left[
\frac{1}{\kappa}
+\zeta D
-\alpha_B(x)
\right]e^y
-C_\eta ,
}
\label{eq:finite_eta_scalar}
\end{equation}
where $C_\eta$ is an integration constant fixed by density conservation and by  the periodicity of the
solution by
\begin{equation}
y(0)=y(L),
\qquad
y'(0)=y'(L),
\qquad
\frac{1}{L}\int_0^L e^{y(x)}\,dx=1.
\label{eq:finite_eta_bc}
\end{equation}
Equation~\eqref{eq:finite_eta_scalar} retains both the viscous stress and the
full concentration-dependent friction $\zeta\phi v$.

The zero-viscosity result derived above follows directly from
Eq.~\eqref{eq:finite_eta_scalar}. Setting $\eta=0$ gives
\begin{equation}
\phi_{\rm 1D}^{(\eta=0)}(x)
=
\frac{C_0}
{\kappa^{-1}+\zeta D-\alpha_B(x)}.
\label{eq:finite_eta_zero_limit}
\end{equation}
For $\alpha_B(x)=-\alpha_B^0f(x)$ this becomes
\begin{equation}
\phi_{\rm 1D}^{(\eta=0)}(x)
=
\frac{\widetilde C_0}{1+Rf(x)},
\qquad
R
=
\frac{\alpha_B^0\kappa}
{1+\kappa\zeta D},
\end{equation}
recovering Eq.~\eqref{eq:rho_general_R}.

For comparison with the interfacial vortex simulations, we use the same
periodic activity pattern employed in the numerical calculation,
\begin{equation}
f(x)
=
1-\frac{1}{2}
\left[
\tanh\!\left(\frac{x-L/4}{W}\right)
-
\tanh\!\left(\frac{x-3L/4}{W}\right)
\right],
\label{eq:finite_eta_vortex_profile}
\end{equation}
with
\begin{equation}
\alpha_B(x)=-\alpha_B^0f(x).
\end{equation}
The active regions therefore lie outside the central passive strip, as in
the two-dimensional vortex simulations.

The distinction between the hydrodynamic screening length and the scale
governing viscous corrections to the density profile is important here.
The screening length
\begin{equation}
\ell_s=\sqrt{\frac{\eta}{\zeta}}
\end{equation}
compares viscous and frictional stresses in the velocity equation.
By contrast, after imposing the steady zero-flux condition, viscosity enters
the density problem through the curvature term
$\eta D\,\partial_x^2\ln\phi$ in
Eq.~\eqref{eq:finite_eta_scalar}. Linearizing this equation about a local
steady scalar balance gives a density-smoothing scale
\begin{equation}
\ell_\phi
\sim
\sqrt{\frac{\eta D}{C_\eta}},
\label{eq:density_viscous_length}
\end{equation}
which is distinct from $\ell_s$.

For the vortex-state parameters, the interface width is $W=16$. At the
simulation value $\eta=1$, the hydrodynamic screening length is
$\ell_s=100$, so that $W<\ell_s$. Thus viscous stresses are indeed important
relative to friction in the momentum balance on the interfacial scale.
Nevertheless, the direct solution of
Eq.~\eqref{eq:finite_eta_scalar} shows that they produce only a small
correction to the steady density redistribution.

\begin{figure}[htpb!]
    \centering
    \includegraphics[
        width=0.72\textwidth
    ]{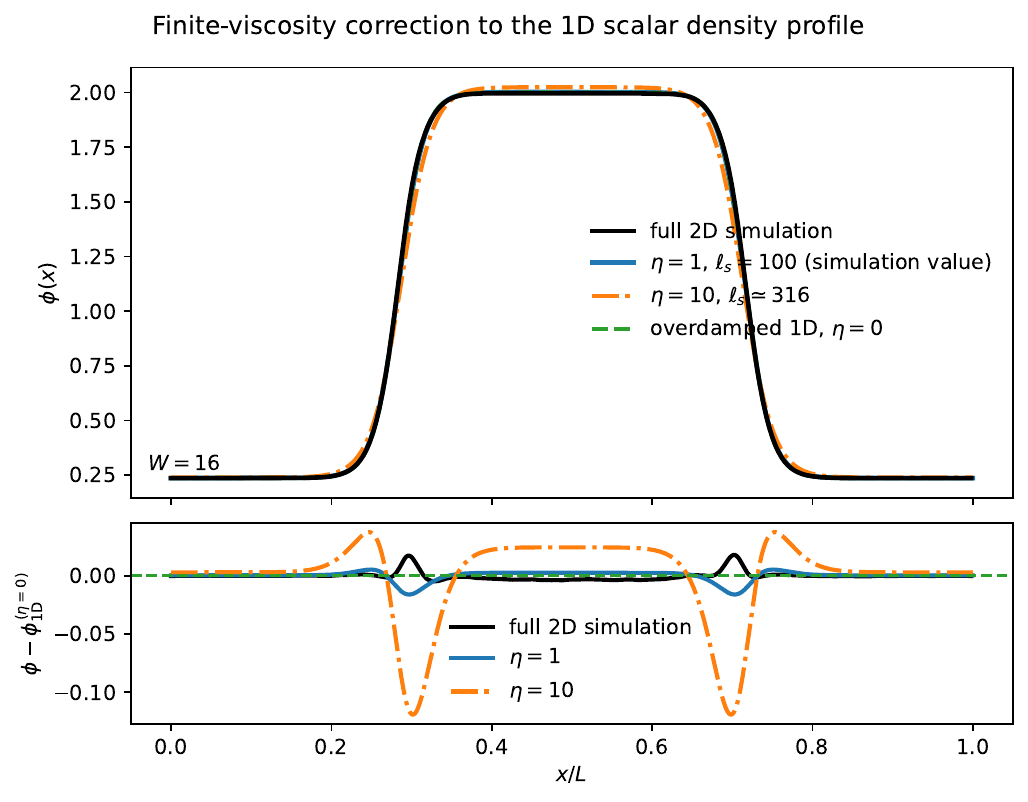}
    \caption{\textcolor{red}{
    Effect of viscosity on the one-dimensional steady density profile.
    Top: the late-time-averaged density profile from the full
    two-dimensional simulation is compared with the finite-viscosity
    one-dimensional solution at $\eta=1$ ($\ell_s=100$), the value used
    in the simulations, the overdamped analytic solution
    $\phi_{\rm 1D}^{(\eta=0)}$, and an additional stress test at
    $\eta=10$ ($\ell_s\simeq316$). The interface width is $W=16$.
    Bottom: residual of each profile relative to the overdamped
    one-dimensional solution,
    $\phi(x)-\phi_{\rm 1D}^{(\eta=0)}(x)$. The overdamped solution itself
    therefore corresponds to the zero line. For $\eta=1$, the viscous
    correction is very small and localized near the interfaces, and the
    full two-dimensional simulation closely follows the same density
    redistribution. Even increasing the viscosity by an order of magnitude
    produces only modest interfacial smoothing while leaving the overall
    redistribution essentially unchanged.
    }}
    \label{fig:finite_eta_scalar}
\end{figure}

Figure~\ref{fig:finite_eta_scalar} compares the finite-viscosity
one-dimensional solution with the overdamped profile and with the
late-time-averaged density profile obtained from the full two-dimensional
simulation. For $\eta=1$, the finite-viscosity and overdamped
one-dimensional profiles are nearly indistinguishable: the density contrast
changes by only $0.13\%$, and the rms difference between them is
$4.4\times10^{-3}$. The corresponding density-smoothing length is
$\ell_\phi\simeq2.23$, well below the imposed interface width $W=16$.
The full two-dimensional density profile exhibits the same redistribution
and closely follows the one-dimensional predictions.

As an additional  test, we also solve
Eq.~\eqref{eq:finite_eta_scalar} for $\eta=10$, an order of magnitude larger
than the value used in the simulations. In this case
$\ell_s\simeq316$ and $\ell_\phi\simeq7.02$. The viscous correction becomes
more visible near the two activity interfaces, but the density contrast
still differs from the overdamped prediction by only approximately $1.2\%$.
We emphasize that $\eta=10$ is included only to test the robustness of the
scalar approximation; all two-dimensional simulations reported in the
manuscript use $\eta=1$.

Thus, although the interface lies in the hydrodynamically
viscosity-dominated regime $W<\ell_s$, this does not imply a large viscous
correction to the steady density redistribution. The hydrodynamic screening
length controls the crossover between viscous and frictional stresses in
the velocity field, whereas the finite-viscosity correction to the scalar
density profile is governed by the much shorter scale $\ell_\phi$. For the
parameters of the vortex state, the finite-viscosity solution therefore
remains quantitatively close to the overdamped scalar profile used in the
analytical treatment and to the density profile obtained from the full
two-dimensional dynamics.

\REVEND

\section{2D Simulations}

\REVSTART

\begin{figure}[htpb!]
    \centering
    \includegraphics[width=0.9\textwidth]{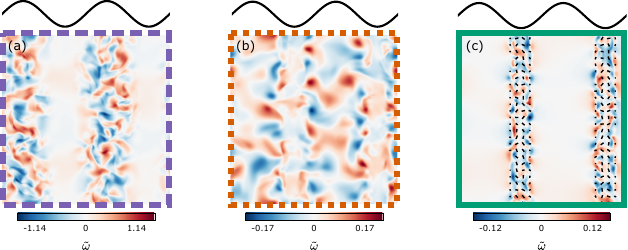}
    \caption{\textcolor{red}{
    \textbf{Expanded views of the representative states in Fig.~3(a) of the main text.}
    \textbf{(a--c)} $\kappa=0.5$, $5$, and $20$, respectively.
    In (c), the director field is overlaid on the vorticity field, with director
    stick length weighted by the local nematic-order magnitude.
    }}
    \label{fig:si_fig3_expanded}
\end{figure}

\REVEND

The numerical results in the main text are obtained from two-dimensional numerical solutions of Eqs.~(1)--(3) in the main text, using pseudospectral solvers based on the \texttt{cuPSS} library~\cite{caballero2024cupss}. The equations are solved on a regular, periodic square grid of size $L\times L$; details of the numerical implementation can be found at \cite{krommydas2026compressible}. Unless stated otherwise, the numerical solutions are obtained on a $512\times512$ grid with periodic boundary conditions, spacing $dx=dy=1$, and solver time step $dt=10^{-3}$. Each run is evolved for at least $5\times10^5$ iterations, corresponding to a total solver time $t_{\rm tot}=500$, and configurations are saved every $5000$ steps. Throughout, we fix $\rho_0=0.7$, $\rhoIN=1$, and hence $\phiIN=\rhoIN/\rho_0=1.43$, together with $K=r_0=0.2$, $D=\eta=\lambda_N=\gamma=u_0=1$, and $\zeta=10^{-4}$, while $\kappa$, the activity parameters $\alpha$ and $\alpha_B$, and the parameters of the imposed activity profile are varied as indicated in the main text and figure captions. In particular, the activity patterns considered are either sinusoidal, with wavelength $\lambda$, or square-pulse profiles regularized by hyperbolic tangents, with interfacial width $W$.

Lengths are reported in units of the bare liquid-crystal coherence length $\ell_c=\sqrt{K/r_0}=1$, times in units of the nematic relaxation time $\tau=\gamma/r_0=5$, and energies in units of the elastic constant $K=0.2$. Thus, the solver time step $dt=10^{-3}$ corresponds to the dimensionless time step $\Delta t=dt/\tau=2\times10^{-4}$ used in the main text, and $t_{\rm tot}=500$ corresponds to a dimensionless integration time $t_{\rm tot}/\tau=100$. Likewise, the solver values $D=u_0=1$ correspond to the dimensionless values $D=5$ and $u_0=5$ quoted in the main text. The normalized density is $\phi=\rho/\rho_0$. Vorticity is normalized by the maximum vorticity obtained from the numerical solution of the incompressible analog of Eqs.~(1)--(3), with all other numerical parameters unchanged. For panels (b)--(c), the solver parameters are $\alpha=-1.0$ and $\alpha_B=-0.7$, corresponding to the dimensionless values $\alpha=-5.0$ and $\alpha_B=-3.5$ used in the main text.

\REVSTART

For reference, Fig.~\ref{fig:si_fig3_expanded} shows enlarged views of the
three representative patterned states used in Fig.~3(a) of the main text.
The enlarged high-compressibility state makes the nematic director field
visible in the low-activity region.

\REVEND

\REVSTART

\subsection{Numerical convergence}

\begin{figure}[htpb!]
    \centering
    \includegraphics[width=\textwidth]{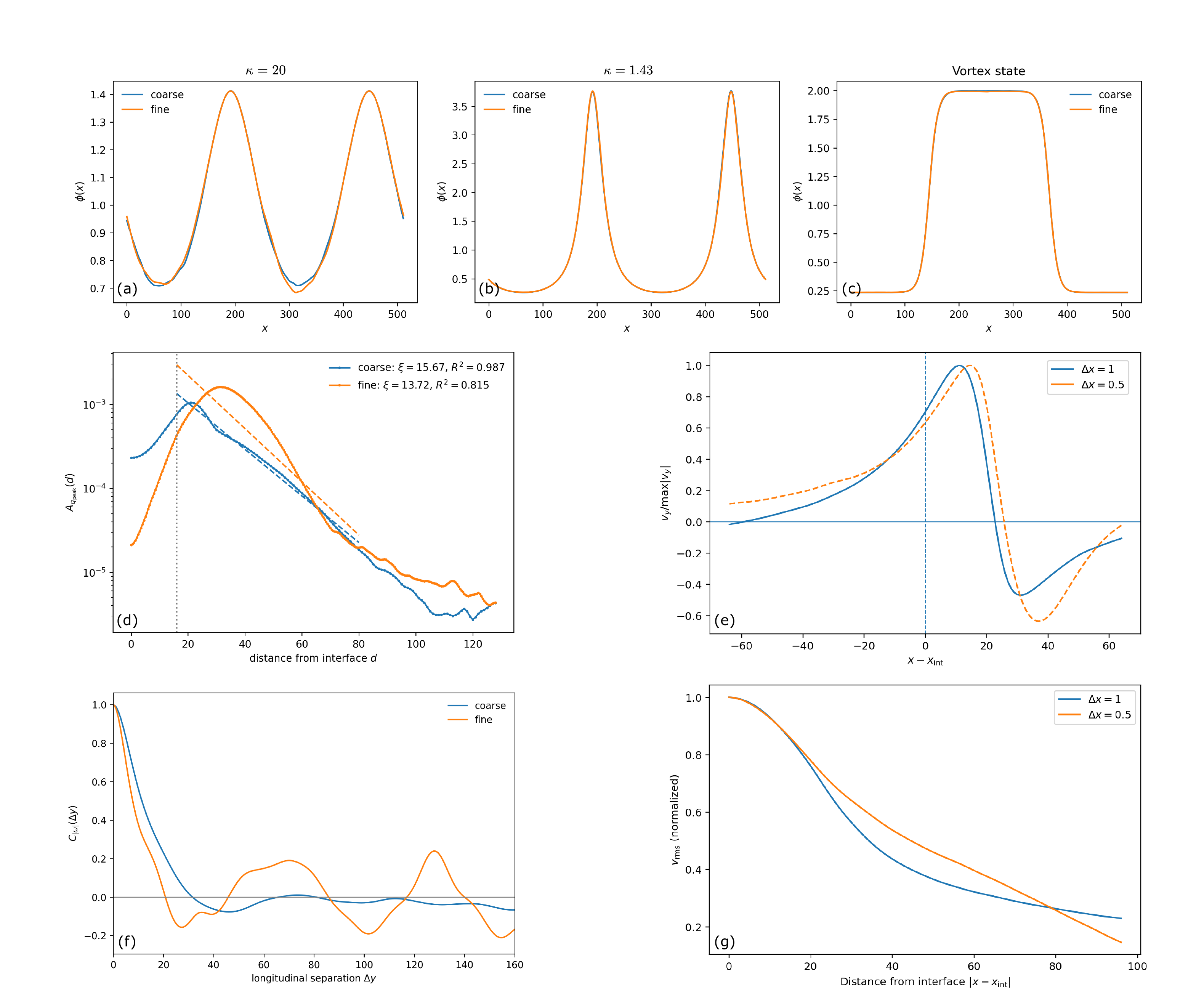}
    \caption{\textcolor{red}{
    \textbf{Spatial-convergence tests.}
    Comparison of $\Delta x=1$ and $\Delta x=0.5$.
    \textbf{(a--c)} Steady density profiles for $\kappa=20$, $\kappa=1.43$,
    and the vortex-chain state.
    \textbf{(d)} Transverse decay of the dominant vortex mode.
    \textbf{(e)} Interface-parallel velocity through a vortex core.
    \textbf{(f)} Longitudinal vorticity autocorrelation.
    \textbf{(g)} RMS flow speed versus distance from the activity interface.
    }}
    \label{fig:si_convergence}
\end{figure}

We tested spatial convergence by repeating representative simulations at
$\Delta x=0.5$, keeping the physical system size and all continuum
parameters fixed. As shown in Fig.~\ref{fig:si_convergence}, the density
profiles and vortex diagnostics are robust to doubling the spatial
resolution. For the matched configurations shown, the first negative
minimum of the longitudinal vorticity autocorrelation occurs at
$\Delta y_{\min}\simeq40$ at both resolutions, corresponding to
$\lambda_v=2\Delta y_{\min}\simeq80$, while the transverse decay length
changes from $15.67$ to $13.72$.

\REVEND

\section{Vortex localization - calculation}

\REVSTART

\subsection{Finite-wavelength instability of the interfacial nematic channel}

To address the spontaneous onset and wavelength selection of the interfacial
nematic texture, we adapt the confined-mode stability analysis of
Valani et al.~\cite{valani2026competing} to the compressible
active-liquid-crystal equations used here. We treat the nematic region
surrounding the activity-induced isotropic--nematic interface as a soft
channel of transverse width of order $W$, with the director aligned on
average along the interface.

The corresponding confined stability problem follows directly from the
homogeneous nematic stability matrix derived above. 
We denote by $q$ the
wave number parallel to the interface and by $k$ the transverse wave number
set by confinement, so that
\begin{equation}
\cos 2\varphi=\frac{q^2-k^2}{q^2+k^2},
\qquad
\sin 2\varphi=\frac{2qk}{q^2+k^2}.
\label{eq:channel_wavevector}
\end{equation}
The full confined problem is obtained by evaluating
Eq.~\eqref{eq:nematic_matrix_dynamics} with the matrix elements in
Eq.~\eqref{eq:nematic_matrix_elements} at total wave number
$\sqrt{q^2+k^2}$ and angle $\varphi$. For $qk\neq0$, concentration and director fluctuations are
coupled, and the onset is determined by the largest eigenvalue of this
$2\times2$ matrix.

The origin of the finite-wavelength mode is particularly transparent in a
director-only reduction in which density feedback is neglected. For
extensile activity, $\alpha=-|\alpha|$, the orientational growth rate then
becomes
\begin{equation}
\sigma_\theta(q)
=
\frac{|\alpha|}
{2\left[\zeta+\eta(q^2+k^2)\right]}
\left[
(\lambda_N+1)q^2+(\lambda_N-1)k^2
\right]
-
\frac{K}{\gamma}(q^2+k^2).
\label{eq:channel_director_growth}
\end{equation}
The neutral activity is therefore
\begin{equation}
|\alpha_c(q)|
=
\frac{
2(K/\gamma)(q^2+k^2)
\left[\zeta+\eta(q^2+k^2)\right]
}{
(\lambda_N+1)q^2+(\lambda_N-1)k^2
},
\label{eq:channel_alpha_c}
\end{equation}
whenever the denominator is positive.

Introducing
\begin{equation}
\beta=\frac{\zeta}{\eta k^2},
\end{equation}
minimization of Eq.~\eqref{eq:channel_alpha_c} gives the finite-wave-number
candidate
\begin{equation}
\frac{q_c^2}{k^2}
=
\frac{
1-\lambda_N+
\sqrt{2\left[2+(\lambda_N+1)\beta\right]}
}{
\lambda_N+1
}.
\label{eq:channel_qc}
\end{equation}
Within this director-only limit, the finite-$q$ minimum merges continuously
with the $q=0$ channel mode at
\begin{equation}
\lambda_N^\ast
=
3+2\beta.
\label{eq:channel_lambda_c}
\end{equation}
Thus, for positive $\lambda_N<\lambda_N^\ast$, the primary orientational
instability is at finite longitudinal wave number. For the parameters used
in our simulations, $\lambda_N=1$ and $W\ll\ell_s$, so
$\beta=(k\ell_s)^{-2}\ll1$ and hence $\lambda_N^\ast\simeq3$; the simulations
therefore lie well within the finite-$q$ regime. More generally, for
order-unity flow-alignment parameters sufficiently below the crossover
$\lambda_N^\ast=3+2\beta$, Eq.~\eqref{eq:channel_qc} gives
$q_c=O(k)$.

For a soft interfacial channel the precise relation between $k$ and $W$
depends on the transverse structure of the activity-induced interface and
its effective boundary conditions. Nevertheless, confinement implies
$k=O(W^{-1})$, and the finite-wave-number instability therefore gives
\begin{equation}
q_c=O(W^{-1}).
\label{eq:channel_qc_scaling}
\end{equation}
A finite longitudinal $q_c$ corresponds to an alternating-sign vortical
mode along the interface. The result therefore supplies the onset and
wavelength-selection mechanism that is absent when the interfacial director
texture is prescribed.

The closed expressions above isolate the orientational mechanism. The full
compressible problem retains the coupled density--director dynamics through
Eq.~\eqref{eq:nematic_matrix_elements}, and hence can shift the quantitative
threshold and selected wave number. We therefore use
Eqs.~\eqref{eq:channel_qc}--\eqref{eq:channel_qc_scaling} to identify the
confinement-controlled finite-wavelength mechanism rather than as an exact
prediction of the nonlinear vortex-chain wavelength. Related recent work by
de Graaf Sousa et al.~\cite{degraafsousa2026spontaneous} identifies
confinement-selected vortex-forming instabilities through a complementary
linear-stability analysis. The calculation below addresses the complementary
post-instability problem: given a developed interfacial nematic modulation,
it determines the resulting velocity, vorticity, and localization away from
the interface.

\REVEND

\subsection{General interfacial solution for an arbitrary active-strength profile}

\subsubsection{Nematic pattern at the active/passive interface }

\begin{figure}[htpb!]
    \centering
    \includegraphics[width=1.0\textwidth]{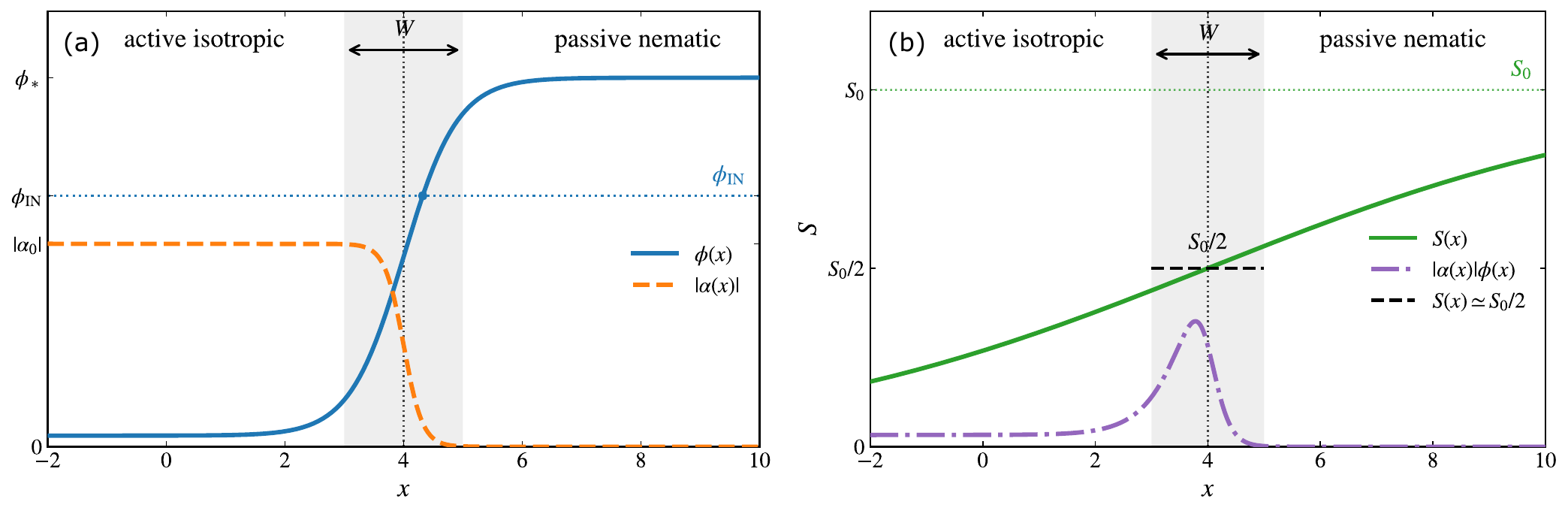}
    \caption{
    \textbf{Schematic profiles across the active/passive interface} centered at $x=x_0$.
    \textbf{(a)} Concentration $\phi(x)$ and activity magnitude $|\alpha(x)|$.
    The active isotropic region has large activity and a concentration close to
    zero, whereas the passive nematic region has vanishing activity and a
    concentration $\phi_\ast \gg\phi_{\mathrm{IN}}$. The concentration crosses
    the isotropic--nematic transition value $\phi_{\mathrm{IN}}$ within the
    shaded interfacial region of width $W$.
    \textbf{(b)} Profiles of $|\alpha(x)| \phi(x) $ and the magnitude of nematic order $S(x)$. $S(x)$ varies smoothly from zero in
    the active isotropic region toward its bulk value $S_0$ in the passive
    nematic region. In the limit $\xi\gg W$, its variation across the narrower
    activity interface is neglected and it is approximated by
    $S(x)\simeq S_0/2$ within the shaded region.
    }
    \label{fig:interfacial_profiles}
\end{figure}

We consider a system with a sharp activity profile $\alpha(x)$ centered at $x=0$ that varies over a length scale $W$ between a passive region with zero activity and a region with a finite activity of magnitude $|\alpha_0|$ (see Fig.~\ref{fig:interfacial_profiles}). Extensile activity pushes microtubules out of the active regions and accumulates them in the passive region, where the concentration is increased to a value  $\phi_*>\phi_{IN}$.  We work in the limit in which the nematic correlation length $\xi$ is larger than $W$, resulting in a smooth variation of the order parameter $S(x)$ across the activity interface between an active/isotropic region for $x<0$ and a passive/nematic region for $x>0$. We then assume that within the interfacial region, $|x|\leq W/2$, the magnitude of the nematic order parameter can be approximated as
\begin{equation}
    S(x)=\frac{S_0}{2}\;,\qquad 
S_0=\sqrt{
\frac{2r_0\left(\phi_\ast-\phiIN\right)}
{u_0\phi_\ast}
}\;.
\label{eq:S0_interface}
\end{equation}
\REVSTART
Having established that the interfacial nematic channel admits a
finite-wavelength instability, we now characterize the flow generated once
such a mode has developed. We represent the leading interfacial director
modulation by a director that is on average aligned with the $y$--direction
and has a small bend modulation along the interface, so that
\REVEND
\begin{equation}
\mathbf{n}\simeq\left(\theta(y),1\right)\;, \qquad
\theta(y) = \theta_0 \sin(qy)\;,
\label{eq:bend}
\end{equation}
where $q = 2\pi m/L$ is a wavenumber, with $m$ an integer,  and $L$ the system size along the $y$--direction.
$\theta_0$ is a small angle (in radians) controlling the bend amplitude.

\REVSTART
The longitudinal wave number $q$ is now identified with the
confinement-selected mode discussed above. For the soft interfacial channel,
$q=q_c=O(W^{-1})$, up to an order-unity factor determined by the transverse
profile and effective boundary conditions.
To obtain a closed-form post-instability flow solution, we linearize the
friction term as $\zeta\phi \approx \zeta$. This approximation is used only
in the analytical calculation below; the full numerical simulations retain
the spatially varying drag $\zeta\phi\mathbf v$ throughout. The Stokes
equation is then given by
\REVEND
\begin{equation}
\zeta v_i-\eta \nabla^2 v_i
=f_i\;.
\label{eq:stokes-bend}
\end{equation}
where flow is driven by active deviatoric forces and passive pressure gradients described by the force density
\begin{equation}
    f_i=\partial_j\big[\alpha(x)\phi(x)\,Q_{ij}\big]
-
\partial_i\left[\left(\frac{1}{\kappa}-\alpha_B(x)\right)\phi(x)\right]\;.
\label{eq:fi}
\end{equation}
Our goal is to solve this equation for the texture profile given by Eqs.~\eqref{eq:S0_interface} and \eqref{eq:bend}, with vanishing velocity away from the active/passive interface, i.e.,  $\lim_{|x|\gg W/2} v_i(x,y) = 0$,
and periodic boundary conditions in $y$.
Clearly vortical flows 
are sourced only by the deviatoric active forcing, corresponding to the first term on the right hand side of Eq.~\eqref{eq:fi}.

For compactness we introduce an effective activity
\begin{equation}
\tilde{\alpha}(x)=\alpha(x)\phi(x)\;.
\end{equation}
This effective activity is appreciable only in the region $|x|<W/2$ and essentially vanishes outside the interfacial region (see Fig.~\ref{fig:interfacial_profiles}).
Using the bend ansatz formulated in the previous section, and keeping terms
to leading order in the bend amplitude $\theta_0$,
we find that in the interfacial region $|x|<W/2$ we can write 
\begin{align}
    &Q_{xx}=-Q_{yy}\simeq-\frac{S_0}{4}\notag\\
    &Q_{xy}\simeq\frac{S_0}{2}\theta(y)\;.
\end{align}
The  force density $f_i$ in the interfacial region $|x|<W/2$ is then given by
\begin{align}    &f_x=-\tilde{\alpha}'\frac{S_0}{4}+\tilde{\alpha}\frac{S_0}{2}\theta_0q\cos(qy)-\kappa^{-1}\phi'+\tilde{\alpha}'
\label{eq:fa-x}\\
&f_y=\tilde{\alpha}'\frac{S_0}{2}\theta_0\sin(qy)
\label{eq:fa-y}
\end{align}
where the prime denotes a derivative wrt $x$.

Given the form of the force density $\mathbf{f}$, we seek solutions of the form
\begin{align}
    &v_x(x,y)=a_x(x)+b_x(x)\cos(qy)\notag\;,\\
     &v_y(x,y)=b_y(x)\sin(qy)\;.
     \label{eq:v-ansatz}
\end{align}
We insert this ansatz into Eq.~\eqref{eq:stokes-bend}, use Eqs.~\eqref{eq:fa-x} and \eqref{eq:fa-y} for the driving force, and equate to zero the coefficient of each harmonic component to obtain a set of ODEs for the coefficients, given by
\begin{align}
&\eta a_x''(x)-\zeta a_x(x)
=
-\tilde{\alpha}'\left(1-S_0/4\right)
+\kappa^{-1}\phi'\;,
\label{eq:ax-ODE-bend}\\
&\eta b_x''(x)-(\eta q^2+\zeta)b_x(x)
=
-\tilde{\alpha}\frac{S_0}{2}\theta_0 q\;,
\label{eq:bx-ODE-bend}\\
&\eta b_y''(x)-(\eta q^2+\zeta)b_y(x)
=
-\tilde{\alpha}'\frac{S_0}{2}\theta_0\;.
\label{eq:by-ODE-bend}
\end{align}

The equations for the first harmonics $b_x$ and $b_y$ contain the same linear operator,
\begin{equation}
\mathcal{L} \equiv
\frac{d^2}{dx^2} - 
\left(q^2 + \frac{\zeta}{\eta}\right)
=
\frac{d^2}{dx^2} - \mu_q^2,
\qquad
\mu_q \equiv \sqrt{q^2 + \frac{\zeta}{\eta}}
=
\sqrt{q^2+\ell_s^{-2}},
\end{equation}
where $\ell_s=\sqrt{\eta/(\zeta)}$ is the flow screening length. 
The solution can be written in terms of the Green's
function $G(x,x')$ defined by
\begin{equation}
\left(\frac{d^2}{dx^2} - \mu_q^2\right)G(x,x')
= \delta(x-x'),
\qquad
\lim_{|x-x'|\to\infty}G(x,x')=0.
\end{equation}
The solution is
\begin{equation}
\boxed{
G(x,x') = -\,\frac{1}{2\mu_q}\,e^{-\mu_q|x-x'|}.
}
\end{equation}
The solutions for the coefficients of the harmonics in the velocity expansion are then given by

\begin{align}
&a_x(x)
=
\frac{\ell_s}{2\eta}
\int_{-\infty}^{\infty} dx'\,
e^{-|x-x'|/\ell_s}
\left[
\left(1-S_0/4\right)\tilde{\alpha}'(x')
-\kappa^{-1}\phi'(x')
\right]\;,\\
&b_x(x)=
\frac{\theta_0 q S_0}{4\eta\mu_q}
\int_{-\infty}^{\infty} dx'\,
e^{-\mu_q|x-x'|}\,
\tilde{\alpha}(x')\;,\\
&b_y(x)=
\frac{\theta_0 S_0}{4\eta\mu_q}
\int_{-\infty}^{\infty} dx'\,
e^{-\mu_q|x-x'|}\,
\tilde{\alpha}'(x')\;.
\label{eq:a-solution-bend}
\end{align}

The modulated flow is screened by the
decay length $\mu_q^{-1}$, which is set by both substrate friction and the
wavenumber $q$. The vorticity, $\omega(x,y)
=
\partial_x v_y(x,y) - \partial_y v_x(x,y)$,
takes a particularly simple form, given by

\begin{align}
\boxed{
\omega(x,y)=
\big[b_y'(x) + q\,b_x(x)\big]\sin(qy)\;.
}
\label{eq:omega}
\end{align}

where in the main text we have used $A(q,x)= b_y'(x) + q\,b_x(x)$.

\REVSTART
Inhomogeneities in the effective activity $\tilde{\alpha}(x)$ along the
$x$ direction generate a steady array of alternating vortices localized
near the interface. The longitudinal size of the vortices is controlled by
the wave number $q$ of the interfacial mode, while their transverse
localization is controlled by $\mu_q^{-1}$. The confined stability
calculation above gives the scaling $q_c=O(W^{-1})$, so the longitudinal
modulation wavelength $2\pi/q_c$ is of order $W$, and the characteristic
vortex dimension along the interface is likewise $O(W)$. The transverse decay length is
\[
\mu_{q_c}^{-1}
=
\left(
q_c^2+\frac{\zeta}{\eta}
\right)^{-1/2}.
\]
For weak hydrodynamic screening on the interfacial scale,
$\ell_s\gg W$, this is also of order $W$. Thus the stability calculation
and the post-instability flow solution together predict an alternating
vortex chain whose characteristic dimensions are controlled by the
interfacial width, up to order-unity factors determined by the soft-channel
profile and effective boundary conditions.
\REVEND
In the following section we present the explicit solution for a specific activity profile.

\subsection{Hyperbolic-tangent step activity profile}

\begin{figure*}[htpb!]
    \centering
\includegraphics[width=0.55\textwidth]{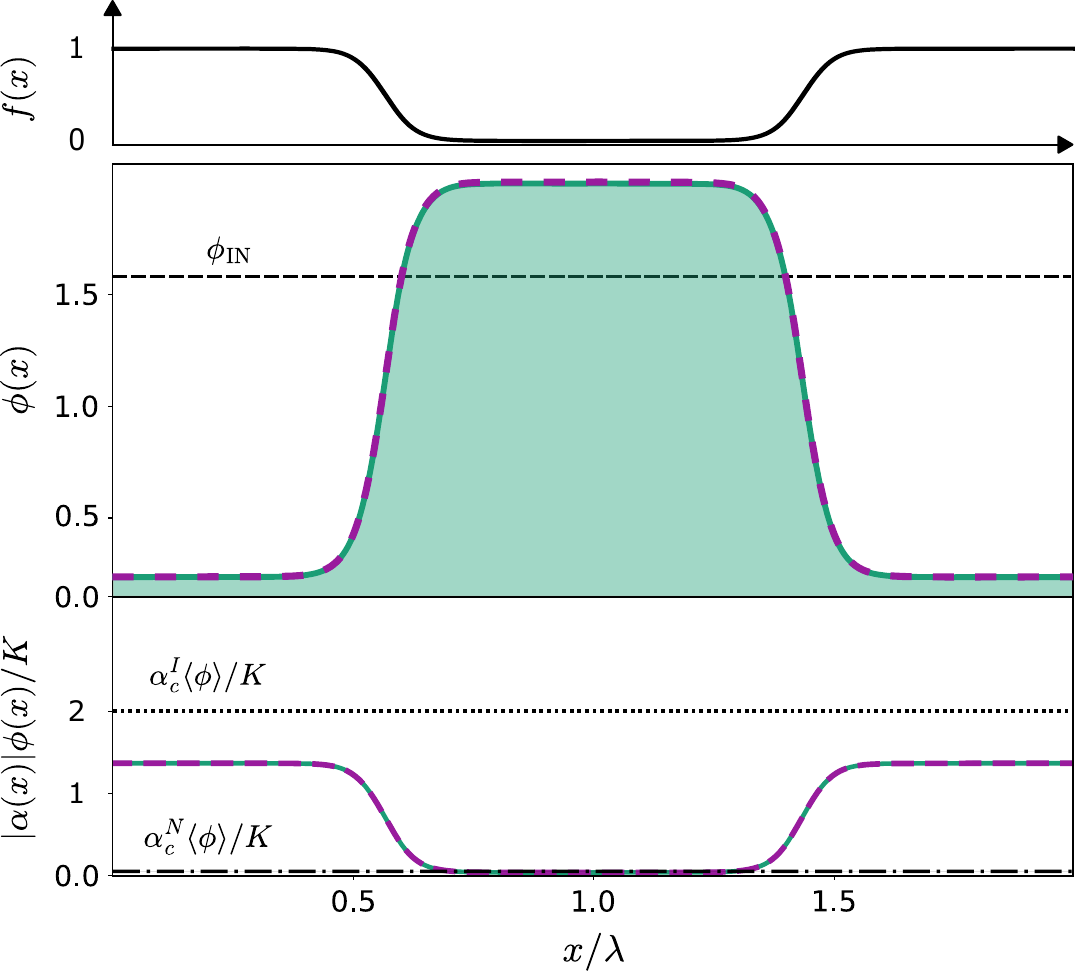}
    \caption{
\textbf{Active-strength profile of the interfacial vortex state.}
Top: Imposed dimensionless activity profile $f(x)$.
Middle: Corresponding normalized density profile $\phi(x)$ for $\kappa=10$, obtained from the full two-dimensional numerical solution (solid green line) and the scalar model (dashed magenta line). The two profiles overlap throughout the domain. The horizontal dashed line indicates the isotropic--nematic transition density, $\phiIN=1.43$.
Bottom: Corresponding dimensionless effective active-strength profile $|\alpha(x)|\phi(x)/K$, obtained from the full two-dimensional numerical solution (solid green line) and the scalar model (dashed magenta line). The two profiles again overlap throughout the domain. The horizontal dotted and dash-dotted lines indicate the thresholds for the onset of active turbulence in the isotropic and nematic states, $\alpha_c^I\langle\phi\rangle/K$ and $\alpha_c^N\langle\phi\rangle/K$, respectively, with $\langle\phi\rangle=1$.
}
    \label{fig:active_strength_profile}
\end{figure*}

The numerical solutions shown in Fig.~3(c) of the main text are carried out with periodic boundary conditions along the $x$ direction, which require a hyperbolic-tangent activity profile given by
\begin{equation}
\alpha(x)=
-\frac{|\alpha|}{2}\left[
\tanh\!\left(\frac{x+\tfrac{L}{4}}{W}\right)
-
\tanh\!\left(\frac{x-\tfrac{L}{4}}{W}\right)
\right],
\label{eq:2tanh}
\end{equation}
The resulting density $\phi$ and $|\alpha(x)| \phi(x)$ profiles for both simulations (green) and scalar model (magenta) are shown in Fig.~\ref{fig:active_strength_profile}. The scalar and full hydrodynamic model are in perfect agreement. 

For simplicity here we focus on the behavior at one of the active/passive interfaces, taken to be centered at $x=0$. We therefore consider the following activity profile
\begin{equation}
\alpha(x)= -\frac{|\alpha|}{2}\,\left[1-\tanh\!\left(\frac{x}{W}\right)\right]\;.
\label{eq:alpha-tanh}
\end{equation}
We assume that the corresponding profile of concentration $\phi(x)$ can be obtained using the scalar model. For a symmetric activity profile consistent with periodic boundary conditions along $x$ like the one given in Eq.~\eqref{eq:2tanh}, the steady state density profile is given by Eq.~\eqref{eq:phi-2tanh}. For the single hyperbolic tangent profile given in Eq.~\eqref{eq:alpha-tanh} and a sharp active/passive interface, corresponding to values of $W$ smaller than the other length scales in the problem, we can approximate
\begin{equation}
\phi(x)= \frac{1}{2}\left[1+\tanh\!\left(\frac{x}{W}\right)\right]\;.
\label{eq:phi-tanh}
\end{equation}
The effective activity $\tilde{\alpha}$ is then given by
\begin{equation}
\boxed{
\tilde{\alpha}(x)=
-\frac{|\alpha|}{4}\,
\sech^2\!\left(\frac{x}{W}\right)\;.
}
\label{eq:g-tanh-sech2}
\end{equation}
Clearly the effective activity is finite only in the interfacial region $|x|\leq W/2$ and vanishes rapidly outside this region, as shown in Fig.~\ref{fig:active_strength_profile}.
In this region, the fluid exhibits nematic order, allowing the onset of turbulent flows. For $x<-W/2$, where activity $\alpha$ is large, the density is low and the fluid is isotropic. The effective activity $\tilde{\alpha}$ is below the threshold for active turbulence and the system is quiescent. For $x>W/2$, the density is above $\phiIN$ and the system is nematic, but the activity $\alpha$ is zero, hence again there is no flow.

Note that $\tilde{\alpha}(x)$ is always even about the interface, $\tilde{\alpha}(-x)=\tilde{\alpha}(x)$, while of course its derivative is odd. 
Since the Green's function in the integral expressions for $b_x(x)$ and $b_y(x)$ is itself even, these symmetry properties imply that $b_x(x)$ is even about the interface, whereas $b_y(x)$ is odd. As a result, the vorticity $\omega(x,y)$ is an even function of $x$ and can therefore peak at the interface itself.

We now examine the explicit solution for the interface profile given in Eq.~\eqref{eq:alpha-tanh}, with derivative
\begin{equation}
\tilde{\alpha}'(x)=
\frac{|\alpha|}{2W}\,
\sech^2\!\left(\frac{x}{W}\right)
\tanh\!\left(\frac{x}{W}\right).
\label{eq:gprime-local}
\end{equation}

For the coefficients of the zeroth harmonics, we take
\begin{equation}
\alpha_B(x)=
-\frac{|\alpha_B|}{2}
\left[
1-\tanh\!\left(\frac{x}{W}\right)
\right].
\label{eq:alphaB-tanh}
\end{equation}

Defining
\begin{equation}
K_0(x)=
\int_{-\infty}^{\infty}
e^{-|x-x'|/\ell_s}
\sech^2\!\left(\frac{x'}{W}\right)\,dx',
\label{eq:K0-def}
\end{equation}
we find
\begin{align}
a_x(x)
&=
\frac{\left(S_0-4\right)|\alpha|\ell_s}
{32\eta}\,
K_0'(x)
-
\frac{\ell_s}
{4\eta\kappa W}\,
K_0(x),
\label{eq:ax-tanh}\\
a_y(x)&=0.
\label{eq:ay-tanh}
\end{align}

For the coefficients of the first harmonics we find
\begin{align}
&b_x(x)= -\frac{\theta_0 q S_0 |\alpha|}{16\eta\mu_q}\,K_{1}(x)
\qquad
 K_1(x)=
\int_{-\infty}^{\infty}
e^{-\mu_q|x-x'|}
\sech^2\!\left(\frac{x'}{W}\right)\,dx'.
\label{eq:a-int-sech2}\\
&b_y(x)= -\frac{\theta_0 S_0 |\alpha|}{8\eta\mu_q W}\,K_{2}(x)
\qquad
 K_2(x)=
\int_{-\infty}^{\infty}
e^{-\mu_q|x-x'|}
\sech^2\!\left(\frac{x'}{W}\right)\tanh\!\left(\frac{x'}{W}\right)\,dx'.
\label{eq:b-int-sech2tanh}
\end{align}

To evaluate \(K_{1}(x)\) we split the integral at \(x'=x\), and set \(y=x'/W\), \(u=x/W\), and \(a\equiv \mu_q W\). Then
\begin{align}
K_{1}(x)
&=
W e^{-a u}
\int_{-\infty}^{u}
e^{a y}\sech^2 y\,dy
+
W e^{a u}
\int_{u}^{\infty}
e^{-a y}\sech^2 y\,dy.
\end{align}
Using \(t=(1+\tanh y)/2\), \(dt=\frac12\sech^2 y\,dy\), and returning to the unscaled variables, we find 
\begin{align}
K_{1}(x)
&=
2W\Bigg[
e^{-\mu_qx}
B_{\frac{1}{1+e^{-\,2x/W}}}
\!\left(
1+\frac{\mu_q W}{2},\,1-\frac{\mu_q W}{2}
\right)
+
e^{\mu_qx}
B_{\frac{1}{1+e^{\,2x/W}}}
\!\left(
1+\frac{\mu_q W}{2},\,1-\frac{\mu_q W}{2}
\right)
\Bigg],
\label{eq:Kx0-exact-new}
\end{align}
where \(B_z(p,q)\) is the incomplete beta function. To evaluate $K_2(x)$, we use
\begin{equation}
\sech^2\!\left(\frac{x}{W}\right)
\tanh\!\left(\frac{x}{W}\right)= -\frac{W}{2}\frac{d}{dx}\sech^2\!\left(\frac{x}{W}\right).
\end{equation}
Integration by parts gives
\begin{align}
K_2(x)=
-\frac{W}{2}\,\frac{d}{dx}K_{1}(x),
\end{align}
where the boundary term vanishes because \(\sech^2(x/W)\to 0\) as
\(|x|\to\infty\). Hence 
\begin{align}
K_{2}(x)
&=-
\mu_q W^2\Bigg[
e^{\mu_qx}
B_{\frac{1}{1+e^{\,2x/W}}}
\!\left(
1+\frac{\mu_q W}{2},\,1-\frac{\mu_q W}{2}
\right)
-
e^{-\mu_qx}
B_{\frac{1}{1+e^{-\,2x/W}}}
\!\left(
1+\frac{\mu_q W}{2},\,1-\frac{\mu_q W}{2}
\right)
\Bigg].
\label{eq:dKdx-exact}
\end{align}

\paragraph{Asymptotic behavior.}

Although the exact expressions involve incomplete beta functions, their
large-distance behavior is simple. Because the source profile and the Green's function both decay exponentially away from the interface ($\sech^2(x/W)\sim e^{-2|x|/W}$), the convolution is governed asymptotically by whichever of the two decays more slowly. Thus, for $|x|\gg W$,
\begin{align}
a_x(x)
&\sim
e^{-\min(\ell_s^{-1},\,2/W)|x|},
\\
b_x(x),\,b_y(x)
&\sim
e^{-\min(\mu_q,\,2/W)|x|}.
\end{align}
The vorticity has the same large-distance decay as the first-harmonic
coefficients. In the regime $\mu_q<2/W$, this reduces to the screened tail
\begin{equation}
b_x(x),\,b_y(x)\sim e^{-\mu_q|x|},
\qquad
\mu_q=\sqrt{q^2+\frac{\zeta}{\eta}}.
\end{equation}
\REVSTART
Using the confinement-controlled scaling $q_c=O(W^{-1})$ obtained above
gives
\REVEND
\begin{equation}
\mu_q
\sim
\sqrt{
W^{-2}+\frac{\zeta}{\eta}
}
=
\sqrt{
W^{-2}+\ell_s^{-2}
}.
\end{equation}
The expressions in Eqs.~\eqref{eq:Kx0-exact-new} and
\eqref{eq:dKdx-exact} are exact for the smooth effective-activity profile
$\tilde{\alpha}(x)$ in Eq.~\eqref{eq:g-tanh-sech2}.

The parameters used to produce Fig.~3(c) (numerical solution) and
Fig.~3(d) (analytical solution) of the main text are
$r_0=0.2$, $u_0=1$, $K=0.2$, $\eta=1$, $\zeta=10^{-4}$,
$\rhoIN=1$, and $\rho_0=0.7$, corresponding to $\phiIN=1.43$.
The system size is $L=512$, and the interfacial width is \textcolor{red}{$W=L/32=16$.}

\begin{figure}[htpb!]
    \centering
    \includegraphics[width=0.9\textwidth]{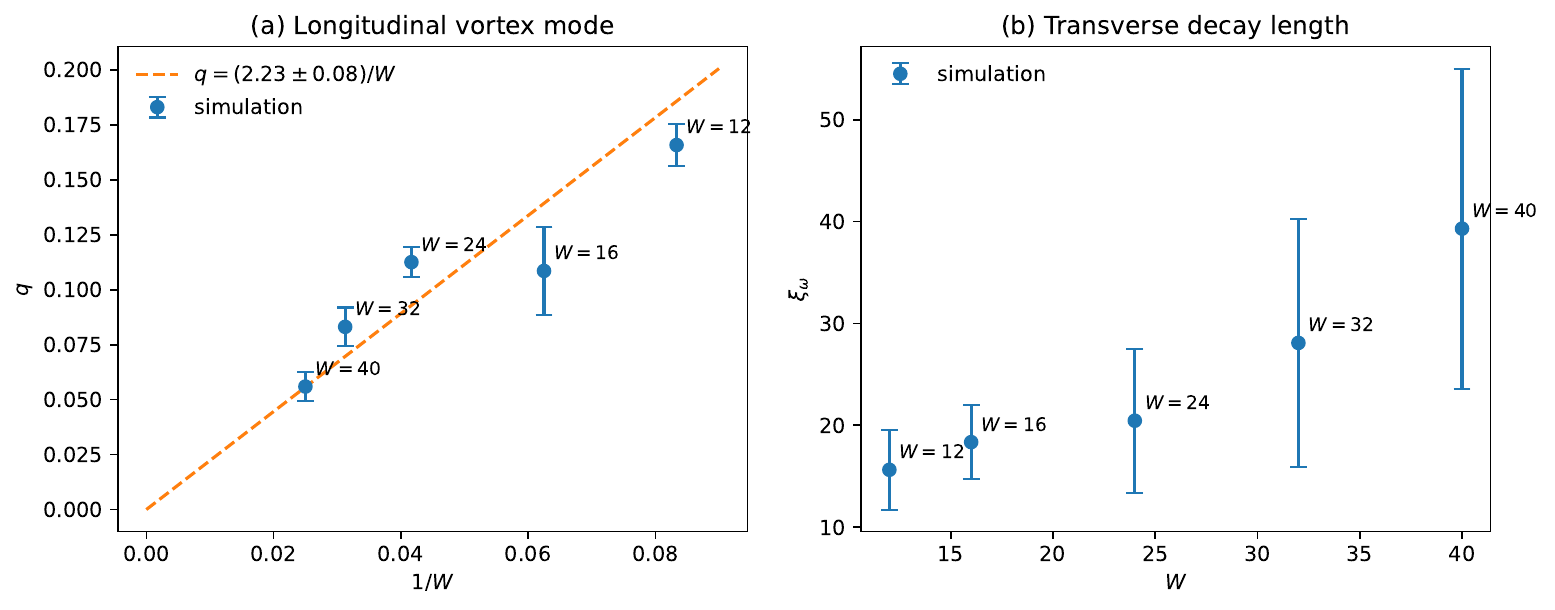}
    \caption{\color{red}
    \textbf{Dependence of the interfacial vortex chain on the interface width.}
    \textbf{(a)} Longitudinal vortex wave number $q$ versus $1/W$.
    The dashed line is the constrained fit $q=(2.23\pm0.08)/W$.
    \textbf{(b)} Transverse decay length $\xi_\omega$ versus $W$.
    Error bars denote the standard deviation over the two interfaces and
    three late-time configurations spanning the final $10^5$ simulation steps.
    }
    \label{fig:si_vortex_width_scaling}
\end{figure}

\REVSTART
We also tested the confinement scaling numerically by varying the interfacial
width over $W=12$--$40$. For each interface, the longitudinal wave number
$q$ is obtained from the first negative minimum of the vorticity
autocorrelation, $q=\pi/\Delta y_{\min}$. As shown in
Fig.~\ref{fig:si_vortex_width_scaling}, the measured wave number is consistent
with the confinement-controlled scaling $q=O(W^{-1})$, while the transverse
decay length increases with $W$.

\REVEND

\section*{Supplemental Movies}

Unless stated otherwise, lengths are reported in units of
$\ell_c=\sqrt{K/r_0}$, times in units of
$\tau=\gamma/r_0$, density as $\phi=\rho/\rho_0$, and
vorticity as the dimensionless quantity $\omega\tau$.

\medskip

\noindent
\textbf{Movie S1.}
Evolution of the normalized density $\phi=\rho/\rho_0$ under a
sinusoidal activity profile of wavelength
$\lambda=256\,\ell_c$, for $\kappa=20$, $\alpha=-5$, and
$\alpha_B=-7.5$.
The movie spans $91\leq t/\tau\leq299$.

\medskip

\noindent
\textbf{Movie S2.}
Evolution of the dimensionless vorticity $\omega\tau$ for the
same parameters as Movie S1.
Red and blue denote opposite signs of the vorticity.
The movie spans $91\leq t/\tau\leq299$.

\medskip

\noindent
\textbf{Movie S3.}
Evolution of the local nematic-order magnitude
$\mathrm{Tr}\,Q^2$ for the same parameters as Movie S1.
The movie spans $91\leq t/\tau\leq299$.

\medskip

\noindent
\textbf{Movie S4.}
Evolution of the normalized density $\phi=\rho/\rho_0$ under a
sinusoidal activity profile of wavelength
$\lambda=256\,\ell_c$, for $\kappa=5$, $\alpha=-5$, and
$\alpha_B=-7.5$.
The movie spans $4\leq t/\tau\leq99$.

\medskip

\noindent
\textbf{Movie S5.}
Evolution of the dimensionless vorticity $\omega\tau$ for the
same parameters as Movie S4.
Red and blue denote opposite signs of the vorticity.
The movie spans $4\leq t/\tau\leq99$.

\medskip

\noindent
\textbf{Movie S6.}
Evolution of the normalized density $\phi=\rho/\rho_0$ under a
sinusoidal activity profile of wavelength
$\lambda=256\,\ell_c$, for $\kappa=1.11$, $\alpha=-5$, and
$\alpha_B=-7.5$.
The movie spans $4\leq t/\tau\leq99$.
For smoother visualization, the rendering was uniformly
motion-interpolated from $8$ to $16$ frames per second without
changing the duration of the movie.

\medskip

\noindent
\textbf{Movie S7.}
Evolution of the dimensionless vorticity $\omega\tau$ for the
same parameters as Movie S6.
Red and blue denote opposite signs of the vorticity.
The movie spans $4\leq t/\tau\leq99$.
For smoother visualization, the rendering was uniformly
motion-interpolated from $8$ to $16$ frames per second without
changing the duration of the movie.

\medskip

\noindent
\textbf{Movie S8.}
Evolution of the dimensionless vorticity $\omega\tau$ under a
sinusoidal activity profile of wavelength
$\lambda=64\,\ell_c$, for $\kappa=10$, $\alpha=-7.5$, and
$\alpha_B=-10$.
Red and blue denote opposite signs of the vorticity.
The movie spans $4\leq t/\tau\leq198$.

\medskip

\noindent
\textbf{Movie S9.}
Evolution of the nematic director field under uniform activity,
visualized using line-integral convolution and colored by the
local density, for $\kappa=1.11$, $\alpha=-7$, and
$\alpha_B=-4.5$.
For this simulation, $L=64\,\ell_c$ and
$\langle\rho\rangle=0.9$.
The movie spans $5.0\leq t/\tau\leq9.8$.

\bibliography{bib}